\documentclass[aps,prl,twocolumn,groupedaddress]{revtex4-2}

\usepackage{amsmath}
\usepackage{amssymb}
\usepackage{amsfonts}
\usepackage[dvipsnames]{xcolor}
\usepackage{hyperref}
\hypersetup{
	pdftex,
	pdfstartview=FitH,
	bookmarksnumbered=true,
	pagebackref,
	colorlinks,
	breaklinks,
	urlcolor=Maroon,
	linkcolor=Maroon,
	citecolor=Bittersweet
}
\usepackage{graphicx}
\DeclareGraphicsExtensions{.pdf,.png,.jpg,.mps}
\usepackage{tikz}
\usetikzlibrary{matrix,decorations.pathmorphing,decorations.markings,arrows,positioning,intersections,mindmap,backgrounds}
\usetikzlibrary{positioning,calc}

\usepackage{booktabs}
\usepackage[caption=false,labelformat=simple]{subfig}

\usepackage{physics}
\usepackage{diagbox}

\usepackage{lmodern}
\usepackage{scalerel,accents,trimclip}
\usepackage{CJK}

\usepackage{shuffle}
\usepackage{cleveref}

\usepackage{MorrisIn,lettrine,Romantik}

\newcommand{\kh}[1]{\hat{\kappa}_{#1}}
\newcommand{\Sz}{S_0}
\newcommand{\Cs}{C_s}

\usepackage{graphicx}

\begin{document}

\begin{CJK*}{UTF8}{gbsn}

    \title{Instanton Structure from 325-Loop 3D $\phi^3$ Perturbation Theory}

    \author{Ian Moult}
    \email{ian.moult@yale.edu}
	\author{Bo Wang (王波)}
    \email{bo.x.wang@yale.edu}
	\affiliation{Department of Physics, Yale University, New Haven, CT 06511}

	\date{\today}

	\begin{abstract}
The high order behavior of perturbation theory provides insight into non-perturbative phenomenon in quantum field theory. While extensively studied in simplified models, high order behavior is challenging to access in more realistic field theories due to the complexity of the standard Feynman diagrammatic expansion. Building on recent progress in global sampling on the moduli space of tropical curves, we introduce an annealed importance sampling framework for tropical field theory. Our approach overcomes the exponential weight-variance barrier, allowing precise numerical calculations to high orders. We apply our technique to compute the perturbative series for the two-point function in the massive three-dimensional $\phi^3$ theory through 325 loops. These coefficients determine the instanton action in $\phi^3$ theory to 0.03\% precision, confirm the collective-coordinate exponent, and extract the one-loop determinant and the instanton fluctuation corrections through four loops, in remarkable agreement with direct semiclassical calculations. Our approach is applicable to a variety of related field theories, facilitating the quantitative exploration of their high loop behavior.
\end{abstract}

	\maketitle

	\end{CJK*}

\lettrine[lines=3]{H}{igh} loop behavior of observables in quantum field theory has long been known to provide insight into non-perturbative phenomena, such as instantons \cite{Bender:1969si,Bender:1973rz,Lipatov:1976ny,Brezin:1976vw,Brezin:1976wa,LeGuillou:1990nq} and renormalons \cite{tHooft:1977xjm,Beneke:1998ui}. The modern theory of resurgence sharpens this connection  \cite{Aniceto:2018bis,Marino:2015yie,Dunne:2012ae,Argyres:2012ka}, opening the possibility of extracting nonperturbative information directly from perturbative data, and understanding the analytic structure of quantum field theory.

Detailed studies of high order behavior of perturbative series have been carried out in quantum mechanics~\cite{Bender:1973rz,Zinn-Justin:2004vcw,Zinn-Justin:2004qzw,Escobar-Ruiz:2015nsa,Giorgini:2021esl}, integrable two-dimensional theories~\cite{Marino:2019eym}, matrix models~\cite{Marino:2007te}, Dyson--Schwinger equations~\cite{Borinsky:2021hnd,Borinsky:2022knn}, and tropical field theory~\cite{Balduf:2025idm}. Renormalon-dominated growth has also been observed in high-order lattice perturbation theory~\cite{Bali:2014fea,Bali:2014sja}. In non-integrable field theories in two and three dimensions, the leading large-order behavior has long served as input for resumming series of six to eight loops~\cite{LeGuillou:1979ixc,Serone:2018gjo}. However, extending these tests to more general interacting field theories remains challenging. Ultraviolet (UV) renormalization and the rapidly increasing complexity of multiloop integrals obstruct access to the high order behavior of perturbative series using standard Feynman diagrammatic techniques.

Recently, there has been progress in understanding the moduli space of Feynman graphs at a given perturbative order \cite{Arkani-Hamed:2023mvg,Arkani-Hamed:2023lbd,Borinsky:2025stz,Salvatori:2025oib}. These techniques open the possibility of extremely high order calculations through the direct sampling of this moduli space, avoiding the need to explicitly construct the complete set of Feynman graphs, which becomes prohibitively expensive at high loop orders. Initial implementations of these ideas from the surfaceology  \cite{Salvatori:2025oib} and tropical field theory \cite{Borinsky:2025stz} perspectives have proven promising.

In this \emph{Letter}, we develop an annealed importance sampling \cite{Neal:2001hwj,Jarzynski:1996oqb} method based on tropical field theory \cite{Borinsky:2025stz,Borinsky:2026sof}. This approach overcomes the exponential weight-variance barrier that stalls direct tropical sampling at large loop orders. We illustrate our technique by computing the perturbative series for the two-point function in the massive $\phi^3$ theory in three dimensions to 325 loops \footnote{This particular order was chosen to match the age of Yale. Higher loops can be achieved as desired.}. Our results provide precision access to the instanton in $\phi^3$ theory, and enable us to extract the instanton fluctuation corrections through four loops. More broadly, they facilitate the practical study of the high order behavior of observables in perturbative quantum field theory.

\vspace{0.8em}
\noindent{\bf Three-dimensional $\phi^3$ field theory.} Massive $\phi^3$ theory in three dimensions offers a particularly clean setting to study the connection between perturbative and non-perturbative physics. It retains the full dynamics of an interacting local quantum field theory, yet its super-renormalizable nature greatly simplifies the UV behavior. In particular, the cubic coupling requires no renormalization, and the massive perturbative expansion avoids the usual running-coupling renormalons. These features allow instanton physics to be probed through perturbative large-order behavior without relying on integrability or a truncation of the field-theoretic dynamics. However, even in this setting, the factorial proliferation of Feynman diagrams makes conventional high-order calculations prohibitive. Analytic series therefore stop at six loops~\cite{Borinsky:2021jdb,Schnetz:2025wtu}, and tropical sampling at twenty~\cite{Borinsky:2025stz}. On the semiclassical side, the large orders of $\phi^3$ theory are known at two loops~\cite{Houghton:1978dt,Mckane:1978me,Malatesta:2017xhp,Giorgini:2024aer}.

We define our Lagrangian as
\begin{equation}\label{eq:action}
S=\int\mathrm{d}^3x\left[\tfrac12(\partial\phi)^2+\tfrac12m_0^2\phi^2+\tfrac{g}{3!}\phi^3+h\phi\right]\,,
\end{equation}
whose imaginary-coupling continuation is associated with Lee--Yang criticality~\cite{Fisher:1978pf,Gracey:2025rnz}. Apart from vacuum graphs, the only primitively UV divergent graph is the one-loop tadpole. We can remove all tadpoles by imposing $\langle\phi\rangle=0$. All 1PI diagrams with two or more external legs are then UV and infrared finite for $m_0>0$. We set $m_0=1$ and expand the self-energy $\Sigma(p)$, defined by the full propagator $1/[1+p^2-\Sigma(p)]$, in $y=g^2/(4\pi)^{3/2}$,
\begin{equation}\label{eq:gamma2}
  \Sigma(p)=\sum_{L\ge1}y^L\left(s_L+d_L\,p^2+u_L\,p^4+\cdots\right),
\end{equation}
where $s_L$, $d_L$, and $u_L$ are the first three Taylor coefficients in $p^2$ of the $L$-loop self-energy, with $s_1=\sqrt\pi/2$ and $s_2=\pi/6$. Each coefficient in Eq.~\eqref{eq:gamma2} is a sum over 1PI two-point graphs $G$ with $L$ loops, $2L$ cubic vertices and $E=3L-1$ internal propagators, 
\begin{align*}
    s_L+d_Lp^2+u_Lp^4+\dots=\sum_{G\in\mathcal G_L}\Sigma_G(p),
\end{align*}
weighted by the symmetry factor $1/|\mathrm{Aut}\,G|$. Introducing Schwinger parameters gives
\begin{align*}
  \Sigma_G(p)=\frac{1}{|\mathrm{Aut}\,G|}\int_0^\infty\!\mathrm{d}^E\alpha\;e^{-\sum_e\alpha_e}\,
  U_G(\alpha)^{-3/2}\,e^{-p^2R_G(\alpha)} .
\end{align*}

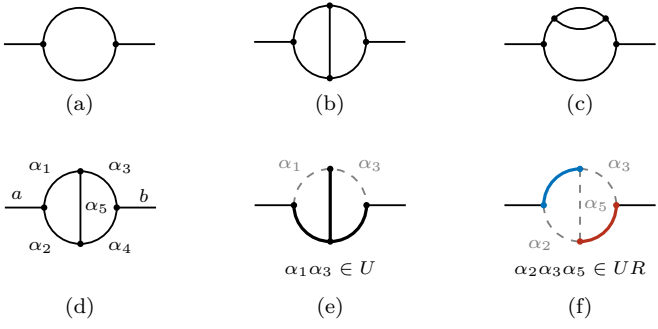
\begin{figure}[t]
  \centering
  \subfloat[]{\label{fig:graphs:a}%
    \begin{tikzpicture}[scale=0.8,line width=0.7pt]   %
      \draw (-1.25,0)--(-0.6,0); \draw (0.6,0)--(1.25,0);
      \draw (0,0) circle (0.6);
      \foreach \p in {(-0.6,0),(0.6,0)} \fill \p circle (1.5pt);
    \end{tikzpicture}}\hfill
  \subfloat[]{\label{fig:graphs:b}%
    \begin{tikzpicture}[scale=0.8,line width=0.7pt]   %
      \draw (-1.25,0)--(-0.6,0); \draw (0.6,0)--(1.25,0);
      \draw (0,0) circle (0.6);
      \draw (0,0.6)--(0,-0.6);
      \foreach \p in {(-0.6,0),(0.6,0),(0,0.6),(0,-0.6)} \fill \p circle (1.5pt);
    \end{tikzpicture}}\hfill
  \subfloat[]{\label{fig:graphs:c}%
    \begin{tikzpicture}[scale=0.8,line width=0.7pt]   %
      \draw (-1.25,0)--(-0.6,0); \draw (0.6,0)--(1.25,0);
      \draw (0,0) circle (0.6);
      \draw (-0.4243,0.4243) arc (225:315:0.6);   %
      \foreach \p in {(-0.6,0),(0.6,0),(-0.4243,0.4243),(0.4243,0.4243)} \fill \p circle (1.5pt);
    \end{tikzpicture}}\\[2pt]
  \subfloat[]{\label{fig:graphs:d}%
    \begin{tikzpicture}[scale=0.8,line width=0.7pt]   %
      \draw (-1.25,0)--(-0.6,0); \draw (0.6,0)--(1.25,0);
      \draw (0,0) circle (0.6); \draw (0,0.6)--(0,-0.6);
      \foreach \p in {(-0.6,0),(0.6,0),(0,0.6),(0,-0.6)} \fill \p circle (1.5pt);
      \node[font=\scriptsize] at (-0.66,0.66) {$\alpha_1$}; \node[font=\scriptsize] at (-0.66,-0.66) {$\alpha_2$};
      \node[font=\scriptsize] at (0.66,0.66) {$\alpha_3$};   \node[font=\scriptsize] at (0.66,-0.66) {$\alpha_4$};
      \node[font=\scriptsize] at (0.27,0) {$\alpha_5$};
      \node[font=\scriptsize] at (-1.05,0.2) {$a$}; \node[font=\scriptsize] at (1.05,0.2) {$b$};
      \node[font=\scriptsize] at (0,-1.05) {\phantom{$\alpha_2\alpha_3\alpha_5$}};
    \end{tikzpicture}}\hfill
  \subfloat[]{\label{fig:graphs:e}%
    \begin{tikzpicture}[scale=0.8,line width=0.7pt]   %
      \draw (-1.25,0)--(-0.6,0); \draw (0.6,0)--(1.25,0);
      \draw[line width=1.2pt] (-0.6,0) arc (180:270:0.6) arc (270:360:0.6);
      \draw[line width=1.2pt] (0,0.6)--(0,-0.6);
      \draw[gray,dashed] (-0.6,0) arc (180:90:0.6) arc (90:0:0.6);
      \foreach \p in {(-0.6,0),(0.6,0),(0,0.6),(0,-0.6)} \fill \p circle (1.5pt);
      \node[font=\scriptsize,gray] at (-0.66,0.66) {$\alpha_1$}; \node[font=\scriptsize,gray] at (0.66,0.66) {$\alpha_3$};
      \node[font=\scriptsize] at (0,-1.05) {$\alpha_1\alpha_3\in U$};
    \end{tikzpicture}}\hfill
  \subfloat[]{\label{fig:graphs:f}%
    \begin{tikzpicture}[scale=0.8,line width=0.7pt]   %
      \draw (-1.25,0)--(-0.6,0); \draw (0.6,0)--(1.25,0);
      \draw[RoyalBlue,line width=1.2pt] (-0.6,0) arc (180:90:0.6);
      \draw[BrickRed,line width=1.2pt] (0,-0.6) arc (270:360:0.6);
      \draw[gray,dashed] (-0.6,0) arc (180:270:0.6); \draw[gray,dashed] (0,0.6) arc (90:0:0.6);
      \draw[gray,dashed] (0,0.6)--(0,-0.6);
      \foreach \p in {(-0.6,0),(0,0.6)} \fill[RoyalBlue] \p circle (1.5pt);
      \foreach \p in {(0.6,0),(0,-0.6)} \fill[BrickRed] \p circle (1.5pt);
      \node[font=\scriptsize,gray] at (-0.66,-0.66) {$\alpha_2$}; \node[font=\scriptsize,gray] at (0.66,0.66) {$\alpha_3$};
      \node[font=\scriptsize,gray] at (0.27,0) {$\alpha_5$};
      \node[font=\scriptsize] at (0,-1.05) {$\alpha_2\alpha_3\alpha_5\in UR$};
    \end{tikzpicture}}
  \caption{\label{fig:graphs}The 1PI two-point graphs through two loops. (a) the one-loop bubble, (b) the two-loop kite, (c) the two-loop self-energy insertion. Each graph has $|\mathrm{Aut}\,G|=2$. (d) The kite with Schwinger parameters $\alpha_e$ and external vertices $a$, $b$. (e) A spanning tree (solid). The two omitted edges (dashed) close the two loops and give the monomial $\alpha_1\alpha_3$ of $U$. (f) A spanning two-forest with one tree containing $a$ (blue) and one containing $b$ (red). Its omitted edges give the monomial $\alpha_2\alpha_3\alpha_5$ of $UR$. The 8 spanning trees and 8 such two-forest give $U=(\alpha_1+\alpha_2)(\alpha_3+\alpha_4)+\alpha_5(\alpha_1+\alpha_2+\alpha_3+\alpha_4)$ and the resistance $R$ of a Wheatstone bridge.}
\end{figure}

The Kirchhoff (first Symanzik) polynomial $U_G$ and the effective resistance $R_G$ are
\begin{align*}
U_G(\alpha)
&=\sum_{T\in\mathcal T_1(G)}
  \prod_{e\notin T}\alpha_e ,
\\
R_G(\alpha)
&=\frac{1}{U_G(\alpha)}
  \sum_{F\in\mathcal T_2^{ab}(G)}
  \prod_{e\notin F}\alpha_e .
\end{align*}
Here $T$ runs over the set $\mathcal T_1(G)$ of spanning trees of $G$, and $F$ over the set $\mathcal T_2^{ab}(G)$ of spanning two-forest pairs of disjoint trees that together contain every vertex, one containing $a$ and the other $b$. Because an $L$-loop graph has $2L$ vertices and $E=3L-1$ edges, each spanning tree $T$ contains $2L-1$ edges and leaves out exactly $L$ loop-closing edges, making each monomial in $U_G$ a product of $L$ parameters. Removing one more edge from $T$ yields a two-forest $F$, which therefore leaves out $L+1$ edges and makes $U_G R_G$ (second Symanzik polynomial~\cite{Bogner:2010kv}) homogeneous of degree $L+1$. Figure~\ref{fig:graphs}(d)--(f) illustrates one term of each for the kite. Physically, $U_G$ is the determinant of the quadratic form in the loop momenta, while $R_G$ is the effective electrical resistance between external vertices $a$ and $b$ when edge $e$ carries resistance $\alpha_e$. Minimizing $\sum_e\alpha_eq_e^2$ over the loop momenta gives $p^2R_G$ by Thomson's principle.

For example, at one loop the only graph is the bubble of Fig.~\ref{fig:graphs}(a), with $|\mathrm{Aut}|=2$
\begin{align*}
    U_a &= \alpha_1+\alpha_2,  \quad R_a = \frac{\alpha_1\alpha_2}{\alpha_1+\alpha_2}.
\end{align*}
We get $s_1={\sqrt\pi}/{2}$, $d_1=-{\sqrt\pi}/{24}$ and $u_1={\sqrt\pi}/{160}$. At two loops there are two graphs, both with $|\mathrm{Aut}|=2$, see Fig.~\ref{fig:graphs}. Expanding the two-loop integrands in $p^2$ reduces the required coefficients to massive vacuum integrals, which can be evaluated using mass derivatives of the three-dimensional sunset integral~\cite{Rajantie:1996np,Davydychev:1992mt}. Together the two graphs give $s_2={\pi}/{6}$, $d_2=-{91\pi}/{1944}$ and $u_2={10537\pi}/{874800}$. Summing these integrals over all graphs defines the coefficients at every order. The number of graphs grows factorially, and the symmetry-weighted count reaches $1.3\times10^{175}$ at 100 loops~\cite{Cvitanovic:1978wc}.

\vspace{0.8em}
\noindent{\bf Annealed importance sampling.} To reach high loop orders despite this factorial proliferation, we build a three-stage Monte Carlo hierarchy that progresses from single-diagram tropical integration~\cite{Borinsky:2020rqs} to global tropical sampling~\cite{Borinsky:2025stz}, and finally to an annealed importance sampling~\cite{Neal:2001hwj,Jarzynski:1996oqb} framework. This enables us to compute perturbative coefficients with relative uncertainties of $3\times10^{-4}$ and $9\times10^{-3}$ at 100 and 325 loops.

\emph{(i) Diagram tropical method.} Let us introduce the tropical method for each diagram first. Borinsky's tropical Monte Carlo method~\cite{Borinsky:2020rqs} integrates a single graph. In each Hepp sector, an ordering of the $\alpha_e$, it replaces $U_G$ by its largest monomial,
\begin{equation}\label{eq:Utr}
  U_G^{\rm tr}(\alpha)=\max_T\prod_{e\notin T}\alpha_e ,
\end{equation}
which can be sampled exactly and leaves a bounded weight $(U_G^{\rm tr}/U_G)^{3/2}\le1$. The bubble shows how this works. It has $U_a=\alpha_1+\alpha_2$ and two sectors. In the sector $\alpha_1>\alpha_2$ one has $U_a^{\rm tr}=\alpha_1$, and writing $\alpha_1=\lambda$, $\alpha_2=\lambda\xi$ with $0<\xi<1$, the $\lambda$ integral of $\Sigma_a(0)$ gives $\Gamma(\frac12)(1+\xi)^{-1/2}=\sqrt\pi(1+\xi)^{-1/2}$. The two sectors give
\begin{align*}
s_1=\sqrt\pi\int_0^1\!\mathrm{d}\xi\,\Bigl(\frac{U_a^{\rm tr}}{U_a}\Bigr)^{3/2}\frac{1}{\sqrt{1+\xi}} =\frac{\sqrt\pi}{2}. 
\end{align*}
The tropical part is uniform in $\xi$ and is sampled exactly. The remaining weight $(1+\xi)^{-2}$ lies between $\frac14$ and 1. For a general graph the $E!$ sectors are drawn edge by edge from a recursion over its $2^E$ edge subsets, and $s_G$ is the exact tropical integral times the average of the weight (App.~\ref{app:tropical}). The weight stays bounded, so its variance is always finite. This accelerates the integration of individual graphs but does not eliminate the cost of listing them.
\begin{figure}[t]
  \centering
  \includegraphics{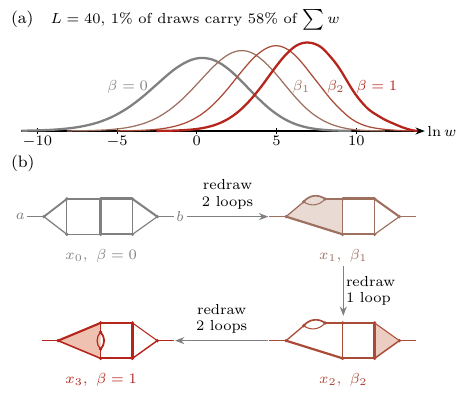}
  \caption{\label{fig:anneal}Annealed tropical sampling. (a)~Distribution of $\ln w$ at $L=40$ under $\pi_\beta$ of Eq.~\eqref{eq:ais}, relative to the mean of the direct draws. Direct draws follow $\beta=0$, but $s_L$ comes from $\beta=1$, where few of them fall. (b)~A chain drawn at four loops. Each step raises $\beta$ and redraws one sub-diagram (shaded) from the tropical measure, with its legs and loop number fixed. Line widths grow with the Schwinger parameters, and colors give $\beta$ as in (a).}
\end{figure}

\emph{(ii) Global tropical method.} Its global extension~\cite{Borinsky:2025stz} instead samples graphs and Schwinger parameters jointly on the moduli space of metric graphs, bypassing explicit graph enumeration. It has reached twenty loops in massive three-dimensional $\phi^3$ theory. It uses the tilted tropical measure 
\begin{equation}
    \mu_\delta=\frac{1}{|\mathrm{Aut}\,G|}\prod_eu_e^{-\delta}\,U^{\rm tr}_G(u)^{-3/2}\prod_e\mathrm{d}u_e
\end{equation}
in cube variables $u_e=1-e^{-\alpha_e}$. Removing the edge with the largest parameter leaves a chain of 1PI beads joined by bridges, so the mass $P_n(L)=\sum_G\int\mu_\delta$ of all $n$-point 1PI graphs obeys
\begin{equation}\label{eq:recursion}
  \omega_\delta(n,L)\,P_n(L)=\tfrac12\,\mathrm{ch}(n,L-1),
\end{equation}
with $\omega_\delta=(1-\delta)(3L-3+n)-3L/2$, a tilt $\delta$, and $\mathrm{ch}(n,l)$ summing over chains of lower-order beads. Exact tropical recursions are not specific to cubic graphs~\cite{Borinsky:2025stz,Balduf:2025idm}. The recursion is particularly simple and positive here, and gives an exact sampler of graphs and parameters at any loop order, with $s_L=P_2(L)\,\mathbb{E}[w]$ and weight $w=(U^{\rm tr}/U)^{3/2}\prod_eu_e^{\delta}\le1$. Here $\mathbb{E}$ averages over $\mu_\delta/P_2(L)$, and $\mu_\delta\,w$ is exactly the integrand of $\Sigma_G(0)$. The tropical measure thus serves as an exactly sampled proposal for the physical integrand, and the momentum dependence follows from the same samples, $d_L=-P_2(L)\,\mathbb{E}[wR_G]$ and $u_L=\frac12P_2(L)\,\mathbb{E}[wR_G^2]$. At higher orders, however, we find that the growing relative variance of the importance weights becomes the dominant computational bottleneck. We introduce the new tilt parameter $\delta$ here compared to the tropical loop equation of Ref.~\cite{Borinsky:2025stz}, which corresponds to $\delta=0$. The factor $\prod_eu_e^{-\delta}$ shifts draws toward small $u_e$, and $w$ carries the inverse factor, so the estimate is exact for any $\delta<1/4$. The tilt $\delta=0.12$ lowers $\mathrm{Var}(\ln w)$ from $0.43L$ to $0.18L$. Direct sampling now works near $L\approx50$, but the variance still increase.

\emph{(iii) Annealed tropical method.} To overcome this variance barrier, we combine global tropical sampling with annealed importance sampling~\cite{Neal:2001hwj,Jarzynski:1996oqb}, replacing direct reweighting by a sequence of intermediate distributions that interpolate between the tropical and physical measures,
\begin{equation}\label{eq:ais}
  \pi_\beta\propto\mu_\delta\,w^\beta,\qquad 0\le\beta\le1 .
\end{equation}
Each run starts from an exact tropical sample. At each step its weight gains a factor $w^{\Delta\beta}$, and a random subtree of the recursion is redrawn from the tropical measure (Fig.~\ref{fig:anneal}(b)) and accepted with probability $\min[1,(w'/w)^\beta]$, satisfy detailed balance. The final weights give an unbiased estimate of $s_L$ for any schedule, and a sweep costs $L^{3.5\text{--}3.8}$. Moreover, at $\beta=1$ the annealed ensemble samples $\pi_1\propto\mu_\delta w$, the normalized integrand of $\Sigma_G(0)$ summed over $L$-loop graphs, allowing ratios of consecutive orders to be extracted from an exact loop-insertion identity, $s_{L+1}/s_L=\mathbb{E}_{\pi_1}[Q]$. Here $Q$ inserts one propagator into the sampled $L$-loop graph, integrates its Schwinger parameter in closed form, and divides by the number of ways the resulting $(L{+}1)$-loop graph is produced. Consecutive ratios thus cost about $10^3$ times less than direct absolute evaluations. Telescoping from the exact $s_2=\pi/6$ gives complete coefficients with full covariance, benchmarked against direct annealed evaluations at high loops, see App.~\ref{app:tropical}. An interactive implementation of all three samplers runs in a web browser~\cite{FanSampler}.

\vspace{0.8em}
\noindent{\bf The instanton and its fluctuations.} Although the coefficients grow factorially, the ratio $s_{L+1}/s_L$ of consecutive coefficients is controlled by a semi-classical solution of the action~\eqref{eq:action}, which we will construct in this section.

\begin{figure}[t]
  \subfloat[]{\label{fig:instanton:a}\includegraphics[height=3.5cm]{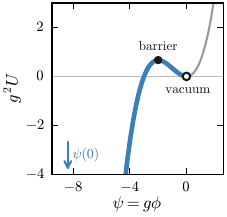}}\hfill
  \subfloat[]{\label{fig:instanton:b}\includegraphics[height=3.3cm]{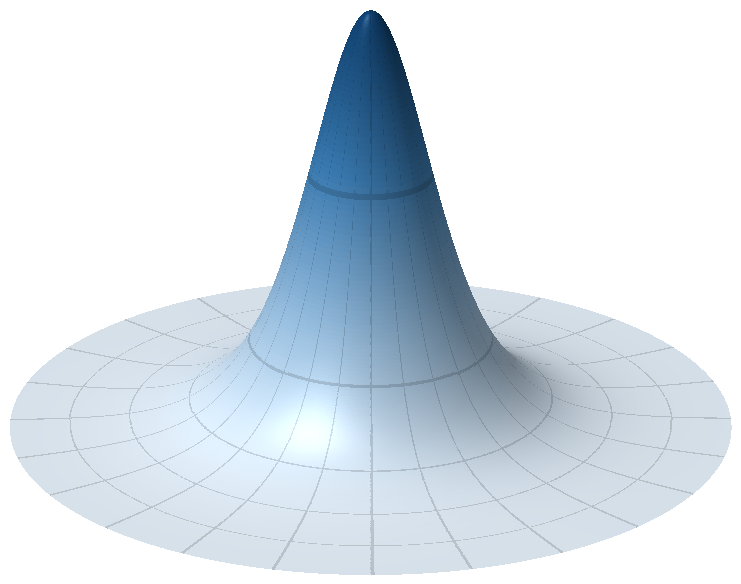}}
  \caption{\label{fig:instanton}(a)~The potential $g^2U=\frac12\psi^2+\frac16\psi^3$, $\psi=g\phi$: the perturbative vacuum $\psi=0$, the barrier top $\psi=-2$, and the unbounded fall to the left. The thick line marks the field values taken by the instanton, $\psi(0)=-8.383$ (arrow). (b)~The instanton, $-\psi$ on the slice $x_3=0$ of Euclidean $\mathbb{R}^3$.}
\end{figure}

For real $g$ the cubic potential is unbounded below (Fig.~\ref{fig:instanton}(a)), and the vacuum $\phi=0$ is only metastable. The Euclidean field equation then has a nontrivial solution that approaches the vacuum at infinity. The maximum principle, the symmetry theorem~\cite{GidasNiNirenberg1979,GidasNiNirenberg1981} and Kwong's uniqueness theorem~\cite{Kwong1989} leave Coleman's bounce~\cite{Coleman:1977py,Callan:1977pt} as the only real one up to translations. With $m_0=1$ this instanton solution $\phi_{\rm cl}=\psi(r)/g$ (Fig.~\ref{fig:instanton}(b)) satisfies
\begin{equation}\label{eq:bounce}
  \psi''+\frac{2}{r}\,\psi'=\psi+\frac12\psi^2,\qquad \psi'(0)=0,\quad \psi(\infty)=0,
\end{equation}
and $\psi(0)=-8.383365909$. Its action is
\begin{equation*}
  S_{\rm cl}=\frac{S_0}{g^2}=\frac{1}{ay},\qquad S_0=174.640946865,\quad a=\frac{(4\pi)^{3/2}}{S_0} .
\end{equation*}
The constant $a=0.2550754836$ converts the action to the coupling $y$. We solve Eq.~\eqref{eq:bounce} numerically by shooting in $\psi(0)$, and the action agrees with the known three-dimensional bounce~\cite{Dine:1992wr}. The field equation and a virial identity give $\int\psi=-\frac32S_0$.

Quantum fluctuations around the instanton follow from $\phi=\phi_{\rm cl}(x-x_0)+\eta$. Since the action is cubic, the expansion terminates,
\begin{equation*}
  S=S_{\rm cl}+\tfrac12\langle\eta,M\eta\rangle+\tfrac{g}{6}\!\int\!\eta^3,\qquad M=-\nabla^2+1+\psi(r),
\end{equation*}
apart from the counterterm $h\int\phi$ with $h=O(g)$. At the instanton it gives $h\int\phi_{\rm cl}=O(1)$, which enters at one loop. The operator $M$ has a single negative mode, at angular momentum $l=0$ with eigenvalue $-2.364$, three zero modes $\partial_\mu\phi_{\rm cl}$ at $l=1$, generated by translations, and a continuum from 1. Along the negative mode the Gaussian integral over $\eta$ diverges, and the instanton contributes an imaginary part, of order $e^{-S_{\rm cl}}$, to the correlation functions.

At one loop and zero momentum the imaginary part of the self-energy $\Sigma(p)$ is a product of four factors. The steepest-descent contour through the negative mode covers half of the Gaussian and gives a factor $\frac12$~\cite{Callan:1977pt}. The zero modes are traded for the position $x_0$, with the Jacobian $(S_{\rm cl}/2\pi)^{3/2}$ since $\|\partial_\mu\phi_{\rm cl}\|^2=S_{\rm cl}$. The remaining modes give a determinant, whose only divergent term, the tadpole $\frac12\mathrm{Tr}\,G_0\psi$ with $G_0=(-\nabla^2+1)^{-1}$, is cancelled exactly by the counterterm $h=-\frac g2G_0(0)+O(g^3)$. This leaves the finite Carleman determinant $\det_2'(M/M_0)$, with the zero modes removed and $M_0=-\nabla^2+1$. A partial-wave Gel'fand--Yaglom evaluation~\cite{Gelfand:1959nq,Dunne:2006ct,Dunne:2005rt} gives $\ln|\det_2'|=-6.13045499$. Determinants of this kind are also computed for three-dimensional bounces in studies of first-order phase transitions~\cite{Ekstedt:2023sqc,Gould:2024chm}. Finally, the two external fields take the instanton value $\int\phi_{\rm cl}=-\frac32S_0/g$,
\begin{align*}
  \mathrm{Im}\,\Sigma(0)&=\frac{1}{2}\Bigl(\frac{S_{\rm cl}}{2\pi}\Bigr)^{3/2}\frac{e^{-S_{\rm cl}}}{|\det_2'|^{1/2}}\Bigl(\int\!\phi_{\rm cl}\Bigr)^2 =\pi C_s\,S_{\rm cl}^{5/2}\,e^{-S_{\rm cl}},\\
  C_s&=\frac{9S_0}{8\pi(2\pi)^{3/2}|\det_2'|^{1/2}}=85.1315052 ,
\end{align*}
where the power $\frac{5}{2}$ counts the three zero modes and the two external fields. At nonzero momentum the external fields carry the Fourier transform $\tilde\psi(p)$ of the profile, amputated by the free propagators.

Beyond one loop the fluctuations correct $\mathrm{Im}\,\Sigma(0)$ in powers of $1/S_{\rm cl}$,
\begin{equation}\label{eq:loops}
  \mathrm{Im}\,\Sigma(0)=\pi C_s\,S_{\rm cl}^{5/2}\,e^{-S_{\rm cl}}\Bigl[1+\sum_{k\ge1}\frac{\hat\kappa_k}{S_{\rm cl}^{k}}\Bigr],
\end{equation}
where $\hat\kappa_k$ is the $(k{+}1)$-loop correction. Because zero-momentum two-point insertions are generated by differentiating with respect to the mass $m_0^2$, an exact chain rule reduces $\hat\kappa_k$ directly to the vacuum fluctuation graphs of $\eta$ in the one-instanton sector. We generated these graphs up to four loops, see App.~\ref{app:instanton} for all details. Our methods reproduce the exactly solvable zero- and one-dimensional versions of the calculation, and their sums do not depend on the gauge. Through four loops we predict
\begin{equation*}
  \hat\kappa_1=-14.36586(3),\enspace \hat\kappa_2=45.4506(5),\enspace \hat\kappa_3=-48.15(40).
\end{equation*}

\begin{figure}[t]
  \centering
  \includegraphics[width=\columnwidth]{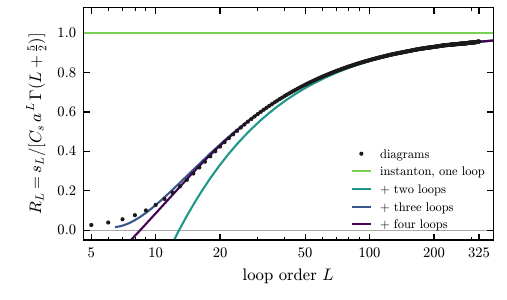}
  \caption{\label{fig:loops}Instanton loops in the complete diagram sum. $R_L$ at every order $L\le325$ against Eq.~\eqref{eq:asym} at one to four loops.}
\end{figure}

\vspace{0.8em}
\noindent{\bf Resurgence from perturbation.} In most field theories, from QED to QCD, perturbative coefficients are what people computed, and instanton or renormalon effects have to be inferred from them~\cite{Beneke:1998ui,Aniceto:2018bis}. Resurgence provides the means, since the divergence of the perturbative series encodes its nonperturbative sectors~\cite{Aniceto:2018bis,Marino:2015yie}. We introduce the Borel transform of $\Sigma(0;y)$,
\begin{align*}
    \Sigma(0;y)=\int_0^\infty\mathrm{d}\zeta\, \frac{e^{-\zeta/y}}{y}\mathcal{B}(\zeta) \, , \quad \mathcal{B}(\zeta)=\sum_Ls_L\frac{\zeta^L}{L!} \, ,
\end{align*}
which has a branch point at $\zeta=1/a$. The form of $\mathcal{B}$ near the branch point is equivalent to the large-order behavior of the coefficients,
\begin{equation}\label{eq:asym}
\begin{aligned}
  s_L&=\Cs\,a^L\,\Gamma(L+b)\,R_L,\\
  R_L&=1+\sum_{k\ge1}\frac{\kh{k}}{(L+b-1)\cdots(L+b-k)} .
\end{aligned}
\end{equation}
Each feature of the singularity is a physical property of a nonperturbative sector. The geometric growth $a^L$, set by the position of the singularity, gives the action, $\Sz=(4\pi)^{3/2}/a$. The exponent $b$ of $\Gamma(L+b)$ counts collective coordinates and external fields, for an instanton $\frac32$ from the three translations and 1 from the two external fields. The normalization $\Cs$ combines the Stokes constant with the one-loop determinant, and the corrections $\kh{k}$ are the fluctuation loops around the saddle.

\begin{table}[t]
\caption{\label{tab:inst}Instanton data predicted by the perturbative coefficients and computed semi-classically. The action and the exponent $b$ of $\Gamma(L+b)$ use no input from the saddle. $\Cs$ uses $a$ and $b$, and each $\kh{k}$ uses $a$, $b$, $\Cs$ and the lower $\kh{k}$ (App.~\ref{app:loopdata}).}
\begin{ruledtabular}
\begin{tabular}{lccc}
 & Instanton & Perturbation & Deviation\\
\hline
Action $\Sz$ & 174.6409 & 174.650(57) & $+0.2\sigma$\\
Exponent $b$ & $5/2$ & 2.51(10) & $+0.1\sigma$\\
\quad at $b=\frac52$ & 174.6409 & 174.635(19) & $-0.3\sigma$\\
Normalization $\Cs$ & 85.1315 & 84.99(30) & $-0.5\sigma$\\
Two loops $\kh1$ & $-14.36586(3)$ & $-14.40(20)$ & $-0.2\sigma$\\
Three loops $\kh2$ & 45.4506(5) & 43.8(1.3) & $-1.3\sigma$\\
Four loops $\kh3$ & $-48.15(40)$ & $-42.3(3.9)$& $+1.5\sigma$\\
\end{tabular}
\end{ruledtabular}
\end{table}

For real coupling, $y>0$, the Laplace contour runs into the cut, and the lateral sums $\Sigma(0;y\pm i0)$ acquire opposite imaginary parts $\pm i\,\mathrm{Im}\,\Sigma(0;y)$. By Eq.~\eqref{eq:asym}, $\mathrm{Im}\,\Sigma(0;y)$ has exactly the form of the instanton contribution, Eq.~\eqref{eq:loops}, with $S_{\rm cl}=1/(ay)$ and the power $\frac52$ replaced by $b$. This is the nonperturbative effect that the perturbative series predicts, and the positive $y$ axis is a Stokes line. Conversely, a dispersion relation in the coupling~\cite{Lipatov:1976ny,Brezin:1976vw,Brezin:1976wa,Mckane:1978me,McKane:2018ocs,Zinn-Justin:2002ecy} returns Eq.~\eqref{eq:asym} from Eq.~\eqref{eq:loops}, $s_L=\pi^{-1}\int_0^\infty\mathrm{d}y\,y^{-L-1}\,\mathrm{Im}\,\Sigma(0;y)$. Its integrand balances $y^{-L-1}$, which favors weak coupling, against $e^{-S_{\rm cl}}$, which favors strong coupling, at $S_{\rm cl}\simeq L$, so the $L$-loop coefficient probes the instanton at the coupling $y_L\simeq1/(aL)$, where its action equals the loop order and the semi-classical expansion applies.

Table~\ref{tab:inst} lists the properties of the singularity read off from the perturbative coefficients related to the instanton. Without any input from the saddle the coefficients give the action and the exponent $b$. The fitted exponent $b=2.51(10)$ tests the collective-coordinate measure of the three-dimensional path integral directly, since each translational zero mode contributes a factor $(S_{\rm cl}/2\pi)^{1/2}\propto1/g$, and the three of them together with the two external fields give $b=\frac{5}{2}$. With the action and $b$ fixed, the coefficients give the strength and the fluctuation loops one by one, through four loops. All agree within $1.5\sigma$, so the nonperturbative effect predicted by perturbation theory is the instanton, see Fig.~\ref{fig:loops}.

For imaginary coupling, $y<0$, the side of the Lee--Yang edge singularity~\cite{Yang:1952be,Lee:1952ig,Fisher:1978pf}, the coefficients alternate in sign and the series is Borel summable. Its critical point, however, is not on this ray. Restoring the bare mass $m_0$ in the dimensionless coupling $y=g^2/(4\pi m_0^2)^{3/2}$, the ray ends at $m_0^2=0$, where $y\to-\infty$. The fluctuations raise the inverse propagator at zero momentum, $m_0^2[1+s_1|y|+O(y^2)]$ at small $|y|$, so we expect it to vanish only at $m_0^2=r_c<0$, beyond the end of the ray. Since $\arg y=\pi-\frac32\arg m_0^2$, continuing $m_0^2$ through zero turns the direction of integration by $3\pi/2$, across the Stokes line, onto the anti-Stokes line $\arg y=-\pi/2$, where $|e^{-S_{\rm cl}}|=1$. A shorter turn reaches the imaginary axis only at complex $m_0^2$. The Lee--Yang point thus lies behind a Stokes barrier. The continuation to criticality acquires the instanton sector with exactly the jump $\Cs$ measured here. Hence, near criticality perturbative and instanton contributions enter on an equal footing, and reaching Lee--Yang criticality from perturbation theory requires more discussions.

\begin{figure}[t]
  \centering
  \includegraphics[width=\columnwidth]{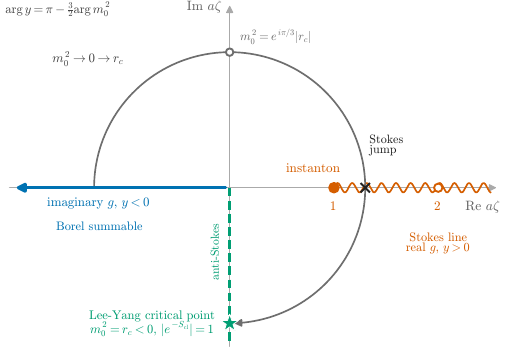}
  \caption{\label{fig:borel}Borel plane of the self-energy, with $\zeta$ in units of $1/a$. The instanton singularities lie on the Stokes line of real $g$, and imaginary coupling is Borel summable. Continuing $m_0^2$ to the Lee--Yang point $r_c<0$ turns the direction of integration by $3\pi/2$ (arc), across the instanton cut. Halfway it passes the imaginary axis at complex $m_0^2$.}
\end{figure}

\vspace{0.8em}
\noindent{\bf Discussion.}  In this \emph{Letter}, we introduced an annealed importance sampling framework for tropical field theory, allowing us ot overcome the exponential weight-variance barrier, and achieve calculations at significantly higher perturbative orders. As an illustration of this framework, we computed the complete perturbative coefficients of the two-point function of three-dimensional $\phi^3$ quantum field theory at every order through 325 loops, at a cost that grows only as a power of the loop order.  At these orders the coefficients are controlled by a single instanton, and they determine its properties one by one. Without input from the saddle they give its action and the exponent of the factorial growth, which counts the translational zero modes and the external fields. With these fixed they give the one-loop normalization and the two-, three- and four-loop fluctuation corrections, all in agreement with an independent semi-classical calculation (Table~\ref{tab:inst}). Summed completely at high order, perturbation theory thus becomes a quantitative measurement of nonperturbative structure.

These results open several directions. Since exact tropical recursions are not specific to cubic graphs~\cite{Borinsky:2025stz,Balduf:2025idm}, the same strategy extends, with mass subtractions, to three-dimensional $\phi^4$ and $O(N)$ models at negative coupling. Renormalizable theories, such as $\phi^4$ in four and $\phi^3$ in six dimensions, require subtracting subdivergences inside the sampler, and fermions and gauge fields require handling numerators and signs. Both extensions would bring within reach theories such as QED and QCD, in which renormalons compete with instantons in the large orders. Finally, the instanton contribution is the input needed to continue the series across the Stokes barrier toward the Lee--Yang point. This establishes a practical route to nonperturbative field configurations through the full diagrammatic expansion.

    \vspace{0.8em}

	\begin{acknowledgments}
	\noindent{\bf Acknowledgments.}
	    The authors would like to thank Michael Saavedra for an inspiring dinner. I.M. thanks Michael Borinsky, Giulio Salvatori, Nima Arkani-Hamed and Carolina Figueiredo for motivating discussions. We acknowledge the use of Claude (Anthropic) and ChatGPT (OpenAI) for code development and implementation, and writing assistance. We take responsibility for all results. I.M. and B.W. are supported by the DOE Early Career Award DE-SC0025581, the Sloan Foundation, and the Simons Collaboration on Confinement and QCD Strings. The authors would also like to thank the Yale Center for Research Computing for computations run on the Bouchet cluster.
	\end{acknowledgments}

	\bibliography{refs}

@article{Arkani-Hamed:2023mvg,
    author = "Arkani-Hamed, N. and Frost, H. and Salvatori, G. and Plamondon, P-G. and Thomas, H.",
    title = "{All loop scattering for all multiplicity}",
    eprint = "2311.09284",
    archivePrefix = "arXiv",
    primaryClass = "hep-th",
    doi = "10.1007/JHEP09(2025)033",
    journal = "JHEP",
    volume = "09",
    pages = "033",
    year = "2025"
}

@article{Arkani-Hamed:2023lbd,
    author = "Arkani-Hamed, N. and Frost, H. and Salvatori, G. and Plamondon, P-G. and Thomas, H.",
    title = "{All loop scattering as a counting problem}",
    eprint = "2309.15913",
    archivePrefix = "arXiv",
    primaryClass = "hep-th",
    doi = "10.1007/JHEP08(2025)194",
    journal = "JHEP",
    volume = "08",
    pages = "194",
    year = "2025"
}

@article{Bender:1969si,
    author = "Bender, Carl M. and Wu, Tai Tsun",
    title = "{Anharmonic oscillator}",
    doi = "10.1103/PhysRev.184.1231",
    journal = "Phys. Rev.",
    volume = "184",
    pages = "1231--1260",
    year = "1969"
}

@article{Bender:1973rz,
    author = "Bender, Carl M. and Wu, T. T.",
    title = "{Anharmonic oscillator. II. A Study of perturbation theory in large order}",
    doi = "10.1103/PhysRevD.7.1620",
    journal = "Phys. Rev. D",
    volume = "7",
    pages = "1620--1636",
    year = "1973"
}

@article{Lipatov:1976ny,
    author = "Lipatov, L. N.",
    title = "{Divergence of the perturbation-theory series and the quasi-classical theory}",
    reportNumber = "LENINGRAD-76-255",
    journal = "Sov. Phys. JETP",
    volume = "45",
    pages = "216--223",
    year = "1977",
    note = "[Zh. Eksp. Teor. Fiz. 72, 411 (1977)]"
}

@article{Brezin:1976vw,
    author = "Br{\'e}zin, E. and Le Guillou, J. C. and Zinn-Justin, Jean",
    title = "{Perturbation Theory at Large Order. I. The $\phi^{2N}$ Interaction}",
    reportNumber = "SACLAY-DPh-T/76-102",
    doi = "10.1103/PhysRevD.15.1544",
    journal = "Phys. Rev. D",
    volume = "15",
    pages = "1544--1557",
    year = "1977"
}

@article{Brezin:1976wa,
    author = "Br{\'e}zin, E. and Le Guillou, J. C. and Zinn-Justin, Jean",
    title = "{Perturbation Theory at Large Order. II. Role of the Vacuum Instability}",
    reportNumber = "SACLAY-DPh-T/76/119",
    doi = "10.1103/PhysRevD.15.1558",
    journal = "Phys. Rev. D",
    volume = "15",
    pages = "1558--1564",
    year = "1977"
}

@article{Coleman:1977py,
    author = "Coleman, Sidney R.",
    title = "{The Fate of the False Vacuum. 1. Semiclassical Theory}",
    reportNumber = "HUTP-77-A004",
    doi = "10.1103/PhysRevD.15.2929",
    journal = "Phys. Rev. D",
    volume = "15",
    pages = "2929--2936",
    year = "1977",
    note = "[Erratum: Phys. Rev. D 16, 1248 (1977)]"
}

@article{Callan:1977pt,
    author = "Callan, Jr., Curtis G. and Coleman, Sidney R.",
    title = "{The Fate of the False Vacuum. 2. First Quantum Corrections}",
    reportNumber = "HUTP-77-A032",
    doi = "10.1103/PhysRevD.16.1762",
    journal = "Phys. Rev. D",
    volume = "16",
    pages = "1762--1768",
    year = "1977"
}

@book{Zinn-Justin:2002ecy,
    author = "Zinn-Justin, Jean",
    title = "{Quantum Field Theory and Critical Phenomena}",
    edition = "4th",
    series = "International Series of Monographs on Physics",
    volume = "113",
    publisher = "Oxford University Press",
    address = "Oxford",
    year = "2002"
}

@book{LeGuillou:1990nq,
    editor = "Le Guillou, J. C. and Zinn-Justin, J.",
    title = "{Large-Order Behaviour of Perturbation Theory}",
    series = "Current Physics--Sources and Comments",
    volume = "7",
    publisher = "North-Holland",
    address = "Amsterdam",
    year = "1990"
}

@article{McKane:2018ocs,
    author = "McKane, Alan J.",
    title = "{Perturbation expansions at large order: Results for scalar field theories revisited}",
    eprint = "1807.00656",
    archivePrefix = "arXiv",
    primaryClass = "hep-th",
    doi = "10.1088/1751-8121/aaf768",
    journal = "J. Phys. A",
    volume = "52",
    number = "5",
    pages = "055401",
    year = "2019"
}

@article{Malatesta:2017xhp,
    author = "Malatesta, Enrico M. and Parisi, Giorgio and Rizzo, Tommaso",
    title = "{Two-loop corrections to large order behavior of $\varphi^4$ theory}",
    eprint = "1704.04458",
    archivePrefix = "arXiv",
    primaryClass = "cond-mat.stat-mech",
    doi = "10.1016/j.nuclphysb.2017.07.011",
    journal = "Nucl. Phys. B",
    volume = "922",
    pages = "293--318",
    year = "2017"
}

@article{Serone:2018gjo,
    author = "Serone, Marco and Spada, Gabriele and Villadoro, Giovanni",
    title = "{$\lambda\phi^4$ theory --- Part I. The symmetric phase beyond NNNNNNNNLO}",
    eprint = "1805.05882",
    archivePrefix = "arXiv",
    primaryClass = "hep-th",
    doi = "10.1007/JHEP08(2018)148",
    journal = "JHEP",
    volume = "2018",
    number = "08",
    pages = "148",
    year = "2018"
}

@article{Marino:2019eym,
    author = "Mari{\~n}o, Marcos and Reis, Tom{\'a}s",
    title = "{Renormalons in integrable field theories}",
    eprint = "1909.12134",
    archivePrefix = "arXiv",
    primaryClass = "hep-th",
    doi = "10.1007/JHEP04(2020)160",
    journal = "JHEP",
    volume = "2020",
    number = "04",
    pages = "160",
    year = "2020"
}

@article{Marino:2007te,
    author = "Mari{\~n}o, Marcos and Schiappa, Ricardo and Weiss, Marlene",
    title = "{Nonperturbative Effects and the Large-Order Behavior of Matrix Models and Topological Strings}",
    eprint = "0711.1954",
    archivePrefix = "arXiv",
    primaryClass = "hep-th",
    reportNumber = "CERN-PH-TH-2007-218",
    doi = "10.4310/CNTP.2008.v2.n2.a3",
    journal = "Commun. Num. Theor. Phys.",
    volume = "2",
    pages = "349--419",
    year = "2008"
}

@article{Bali:2014sja,
    author = "Bali, Gunnar S. and Bauer, Clemens and Pineda, Antonio",
    title = "{Model-independent determination of the gluon condensate in four-dimensional SU(3) gauge theory}",
    eprint = "1403.6477",
    archivePrefix = "arXiv",
    primaryClass = "hep-ph",
    doi = "10.1103/PhysRevLett.113.092001",
    journal = "Phys. Rev. Lett.",
    volume = "113",
    pages = "092001",
    year = "2014"
}

@article{Bali:2014fea,
    author = "Bali, Gunnar S. and Bauer, Clemens and Pineda, Antonio",
    title = "{Perturbative expansion of the plaquette to ${\cal O}(\alpha^{35})$ in four-dimensional SU(3) gauge theory}",
    eprint = "1401.7999",
    archivePrefix = "arXiv",
    primaryClass = "hep-ph",
    doi = "10.1103/PhysRevD.89.054505",
    journal = "Phys. Rev. D",
    volume = "89",
    pages = "054505",
    year = "2014"
}

@article{Borinsky:2021hnd,
    author = "Borinsky, Michael and Dunne, Gerald V. and Meynig, Max",
    title = "{Semiclassical Trans-Series from the Perturbative Hopf-Algebraic Dyson-Schwinger Equations: $\phi^3$ QFT in 6 Dimensions}",
    eprint = "2104.00593",
    archivePrefix = "arXiv",
    primaryClass = "hep-th",
    reportNumber = "Nikhef 2021-010",
    doi = "10.3842/SIGMA.2021.087",
    journal = "SIGMA",
    volume = "17",
    pages = "087",
    year = "2021"
}

@article{Borinsky:2022knn,
    author = "Borinsky, Michael and Broadhurst, David",
    title = "{Resonant resurgent asymptotics from quantum field theory}",
    eprint = "2202.01513",
    archivePrefix = "arXiv",
    primaryClass = "hep-th",
    doi = "10.1016/j.nuclphysb.2022.115861",
    journal = "Nucl. Phys. B",
    volume = "981",
    pages = "115861",
    year = "2022"
}

@article{Borinsky:2020rqs,
    author = "Borinsky, Michael",
    title = "{Tropical Monte Carlo quadrature for Feynman integrals}",
    eprint = "2008.12310",
    archivePrefix = "arXiv",
    primaryClass = "math-ph",
    reportNumber = "Nikhef 2020-027",
    doi = "10.4171/aihpd/158",
    journal = "Ann. Inst. H. Poincare D Comb. Phys. Interact.",
    volume = "10",
    number = "4",
    pages = "635--685",
    year = "2023"
}

@article{Borinsky:2025stz,
    author = "Borinsky, Michael",
    title = "{Tropicalized quantum field theory and global tropical sampling}",
    eprint = "2508.14263",
    archivePrefix = "arXiv",
    primaryClass = "math-ph",
    doi = "10.1007/s00023-026-01695-y",
    journal = "Ann. Henri Poincar\'e",
    year = "2026"
}

@misc{Balduf:2025idm,
    author = "Balduf, Paul-Hermann and Panzer, Erik",
    title = "{Renormalized tropical field theory}",
    eprint = "2512.21091",
    archivePrefix = "arXiv",
    primaryClass = "math-ph",
    month = "12",
    year = "2025"
}

@article{Neal:2001hwj,
    author = "Neal, Radford M.",
    title = "{Annealed importance sampling}",
    doi = "10.1023/A:1008923215028",
    journal = "Stat. Comput.",
    volume = "11",
    number = "2",
    pages = "125--139",
    year = "2001"
}

@article{Jarzynski:1996oqb,
    author = "Jarzynski, C.",
    title = "{Nonequilibrium Equality for Free Energy Differences}",
    eprint = "cond-mat/9610209",
    archivePrefix = "arXiv",
    doi = "10.1103/PhysRevLett.78.2690",
    journal = "Phys. Rev. Lett.",
    volume = "78",
    number = "14",
    pages = "2690--2693",
    year = "1997"
}

@article{GidasNiNirenberg1979,
  author        = {Gidas, B. and Ni, Wei-Ming and Nirenberg, L.},
  title         = {Symmetry and related properties via the maximum principle},
  journal       = {Commun. Math. Phys.},
  volume        = {68},
  pages         = {209--243},
  year          = {1979},
  doi           = {10.1007/BF01221125}
}

@incollection{GidasNiNirenberg1981,
  author        = {Gidas, B. and Ni, Wei-Ming and Nirenberg, L.},
  title         = {Symmetry of positive solutions of nonlinear elliptic equations in {$\mathbf{R}^n$}},
  booktitle     = {Mathematical Analysis and Applications, Part A},
  editor        = {Nachbin, L.},
  series        = {Adv. Math. Suppl. Stud.},
  volume        = {7a},
  publisher     = {Academic Press},
  address       = {New York},
  year          = {1981},
  pages         = {369--402}
}

@article{Kwong1989,
  author        = {Kwong, Man Kam},
  title         = {Uniqueness of positive solutions of {$\Delta u - u + u^p = 0$} in {$\mathbf{R}^n$}},
  journal       = {Arch. Ration. Mech. Anal.},
  volume        = {105},
  pages         = {243--266},
  year          = {1989},
  doi           = {10.1007/BF00251502}
}

@article{Fisher:1978pf,
    author = "Fisher, M. E.",
    title = "{Yang-Lee Edge Singularity and $\phi^3$ Field Theory}",
    doi = "10.1103/PhysRevLett.40.1610",
    journal = "Phys. Rev. Lett.",
    volume = "40",
    pages = "1610--1613",
    year = "1978"
}

@article{Gracey:2025rnz,
    author = "Gracey, J. A.",
    title = "{Six loop critical exponent analysis for Lee-Yang and percolation theory}",
    eprint = "2510.05723",
    archivePrefix = "arXiv",
    primaryClass = "hep-th",
    reportNumber = "LTH 1409",
    doi = "10.1103/kphq-sc3j",
    journal = "Phys. Rev. D",
    volume = "112",
    number = "10",
    pages = "105019",
    year = "2025"
}

@article{Schnetz:2025wtu,
    author = "Schnetz, Oliver",
    title = "{$\phi^3$ theory at six loops}",
    eprint = "2505.15485",
    archivePrefix = "arXiv",
    primaryClass = "hep-th",
    doi = "10.1103/sp5q-km2c",
    journal = "Phys. Rev. D",
    volume = "112",
    number = "1",
    pages = "016028",
    year = "2025"
}

@article{Borinsky:2021jdb,
    author = "Borinsky, M. and Gracey, J. A. and Kompaniets, M. V. and Schnetz, O.",
    title = "{Five-loop renormalization of $\phi^3$ theory with applications to the Lee-Yang edge singularity and percolation theory}",
    eprint = "2103.16224",
    archivePrefix = "arXiv",
    primaryClass = "hep-th",
    reportNumber = "Nikhef-2021-006, LTH-1254",
    doi = "10.1103/PhysRevD.103.116024",
    journal = "Phys. Rev. D",
    volume = "103",
    number = "11",
    pages = "116024",
    year = "2021"
}

@article{Escobar-Ruiz:2015nsa,
    author = "Escobar-Ruiz, M. A. and Shuryak, E. and Turbiner, A. V.",
    title = "{Three-loop Correction to the Instanton Density. I. The Quartic Double Well Potential}",
    eprint = "1501.03993",
    archivePrefix = "arXiv",
    primaryClass = "hep-th",
    doi = "10.1103/PhysRevD.92.025046",
    journal = "Phys. Rev. D",
    volume = "92",
    number = "2",
    pages = "025046",
    year = "2015",
    note = "[Erratum: Phys. Rev. D 92, 089902 (2015)]"
}

@article{Zinn-Justin:2004vcw,
    author = "Zinn-Justin, Jean and Jentschura, Ulrich D.",
    title = "{Multi-instantons and exact results I: Conjectures, WKB expansions, and instanton interactions}",
    eprint = "quant-ph/0501136",
    archivePrefix = "arXiv",
    doi = "10.1016/j.aop.2004.04.004",
    journal = "Annals Phys.",
    volume = "313",
    pages = "197--267",
    year = "2004"
}

@article{Zinn-Justin:2004qzw,
    author = "Zinn-Justin, Jean and Jentschura, Ulrich D.",
    title = "{Multi-instantons and exact results II: Specific cases, higher-order effects, and numerical calculations}",
    eprint = "quant-ph/0501137",
    archivePrefix = "arXiv",
    doi = "10.1016/j.aop.2004.04.003",
    journal = "Annals Phys.",
    volume = "313",
    pages = "269--325",
    year = "2004"
}

@article{Dunne:2006ct,
    author = "Dunne, Gerald V. and Kirsten, Klaus",
    title = "{Functional determinants for radial operators}",
    eprint = "hep-th/0607066",
    archivePrefix = "arXiv",
    doi = "10.1088/0305-4470/39/38/017",
    journal = "J. Phys. A",
    volume = "39",
    pages = "11915--11928",
    year = "2006"
}

@article{Gelfand:1959nq,
    author = "Gel'fand, I. M. and Yaglom, A. M.",
    title = "{Integration in functional spaces and its applications in quantum physics}",
    doi = "10.1063/1.1703636",
    journal = "J. Math. Phys.",
    volume = "1",
    pages = "48--69",
    year = "1960"
}

@article{LeGuillou:1979ixc,
    author = "Le Guillou, J. C. and Zinn-Justin, Jean",
    title = "{Critical Exponents from Field Theory}",
    reportNumber = "SACLAY-DPh-T 79/94",
    doi = "10.1103/PhysRevB.21.3976",
    journal = "Phys. Rev. B",
    volume = "21",
    pages = "3976--3998",
    year = "1980"
}

@article{Beneke:1998ui,
    author = "Beneke, M.",
    title = "{Renormalons}",
    eprint = "hep-ph/9807443",
    archivePrefix = "arXiv",
    reportNumber = "CERN-TH-98-233",
    doi = "10.1016/S0370-1573(98)00130-6",
    journal = "Phys. Rept.",
    volume = "317",
    pages = "1--142",
    year = "1999"
}

@article{Aniceto:2018bis,
    author = "Aniceto, In{\^e}s and Ba{\c{s}}ar, G{\"o}k{\c{c}}e and Schiappa, Ricardo",
    title = "{A Primer on Resurgent Transseries and Their Asymptotics}",
    eprint = "1802.10441",
    archivePrefix = "arXiv",
    primaryClass = "hep-th",
    reportNumber = "NSF-ITP-17-153",
    doi = "10.1016/j.physrep.2019.02.003",
    journal = "Phys. Rept.",
    volume = "809",
    pages = "1--135",
    year = "2019"
}

@article{Cvitanovic:1978wc,
    author = "Cvitanovi{\'c}, Predrag and Lautrup, B. E. and Pearson, Robert B.",
    title = "{The Number and Weights of Feynman Diagrams}",
    reportNumber = "Print-78-0597 (IAS,PRINCETON)",
    doi = "10.1103/PhysRevD.18.1939",
    journal = "Phys. Rev. D",
    volume = "18",
    pages = "1939--1949",
    year = "1978"
}

@book{Marino:2015yie,
    author = "Mari{\~n}o, Marcos",
    title = "{Instantons and Large N}: {An Introduction to Non-Perturbative Methods in Quantum Field Theory}",
    doi = "10.1017/CBO9781107705968",
    isbn = "978-1-107-06852-0, 978-1-316-37154-1",
    publisher = "Cambridge University Press",
    month = "9",
    year = "2015"
}

@article{Giorgini:2024aer,
    author = "Giorgini, Ludovico T. and Jentschura, Ulrich D. and Malatesta, Enrico M. and Rizzo, Tommaso and Zinn-Justin, Jean",
    title = "{Instantons in $\phi^4$ theories: Transseries, virial theorems, and numerical aspects}",
    eprint = "2405.18191",
    archivePrefix = "arXiv",
    primaryClass = "hep-th",
    doi = "10.1103/PhysRevD.110.036003",
    journal = "Phys. Rev. D",
    volume = "110",
    number = "3",
    pages = "036003",
    year = "2024"
}

@misc{FanSampler,
  author        = {Moult, Ian and Wang, Bo},
  title         = {{Tropical Fan Sampler}: interactive diagram, global and annealed tropical sampling},
  howpublished  = {\url{https://boyuuwang-oss.github.io/tropical-fan-sampler/}},
  year          = {2026}
}

@article{Dunne:2012ae,
    author = "Dunne, Gerald V. and {\"U}nsal, Mithat",
    title = "{Resurgence and Trans-series in Quantum Field Theory: The CP(N-1) Model}",
    eprint = "1210.2423",
    archivePrefix = "arXiv",
    primaryClass = "hep-th",
    doi = "10.1007/JHEP11(2012)170",
    journal = "JHEP",
    volume = "2012",
    number = "11",
    pages = "170",
    year = "2012"
}

@article{Argyres:2012ka,
    author = "Argyres, Philip C. and {\"U}nsal, Mithat",
    title = "{The semi-classical expansion and resurgence in gauge theories: new perturbative, instanton, bion, and renormalon effects}",
    eprint = "1206.1890",
    archivePrefix = "arXiv",
    primaryClass = "hep-th",
    reportNumber = "UCHEP-12-08",
    doi = "10.1007/JHEP08(2012)063",
    journal = "JHEP",
    volume = "2012",
    number = "08",
    pages = "063",
    year = "2012"
}

@article{tHooft:1977xjm,
    author = "'t Hooft, Gerard",
    editor = "Zichichi, Antonino",
    title = "{Can We Make Sense Out of Quantum Chromodynamics?}",
    reportNumber = "PRINT-77-0723 (UTRECHT)",
    journal = "Subnucl. Ser.",
    volume = "15",
    pages = "943",
    year = "1979"
}

@article{Rajantie:1996np,
    author = "Rajantie, Arttu K.",
    title = "{Feynman diagrams to three loops in three-dimensional field theory}",
    eprint = "hep-ph/9606216",
    archivePrefix = "arXiv",
    reportNumber = "HU-TFT-96-22",
    doi = "10.1016/S0550-3213(96)00474-9",
    journal = "Nucl. Phys. B",
    volume = "480",
    pages = "729--752",
    year = "1996",
    note = "[Erratum: Nucl. Phys. B 513, 761--762 (1998)]"
}

@article{Davydychev:1992mt,
    author = "Davydychev, Andrei I. and Tausk, J. B.",
    title = "{Two loop selfenergy diagrams with different masses and the momentum expansion}",
    reportNumber = "INLO-PUB-11-92",
    doi = "10.1016/0550-3213(93)90338-P",
    journal = "Nucl. Phys. B",
    volume = "397",
    pages = "123--142",
    year = "1993"
}

@article{Bogner:2010kv,
    author = "Bogner, Christian and Weinzierl, Stefan",
    title = "{Feynman graph polynomials}",
    eprint = "1002.3458",
    archivePrefix = "arXiv",
    primaryClass = "hep-ph",
    doi = "10.1142/S0217751X10049438",
    journal = "Int. J. Mod. Phys. A",
    volume = "25",
    pages = "2585--2618",
    year = "2010"
}

@article{Houghton:1978dt,
    author = "Houghton, A. and Reeve, J. S. and Wallace, D. J.",
    title = "{High-order behavior in $\phi^3$ field theories and the percolation problem}",
    reportNumber = "THEP-77-78-1",
    doi = "10.1103/PhysRevB.17.2956",
    journal = "Phys. Rev. B",
    volume = "17",
    pages = "2956--2964",
    year = "1978"
}

@article{Dine:1992wr,
    author = "Dine, Michael and Leigh, Robert G. and Huet, Patrick Y. and Linde, Andrei D. and Linde, Dmitri A.",
    title = "{Towards the theory of the electroweak phase transition}",
    eprint = "hep-ph/9203203",
    archivePrefix = "arXiv",
    reportNumber = "SLAC-PUB-5741, SCIPP-92-07, SU-ITP-92-7",
    doi = "10.1103/PhysRevD.46.550",
    journal = "Phys. Rev. D",
    volume = "46",
    pages = "550--571",
    year = "1992"
}

@article{Dunne:2005rt,
    author = "Dunne, Gerald V. and Min, Hyunsoo",
    title = "{Beyond the thin-wall approximation: Precise numerical computation of prefactors in false vacuum decay}",
    eprint = "hep-th/0511156",
    archivePrefix = "arXiv",
    doi = "10.1103/PhysRevD.72.125004",
    journal = "Phys. Rev. D",
    volume = "72",
    pages = "125004",
    year = "2005"
}

@article{Ekstedt:2023sqc,
    author = "Ekstedt, Andreas and Gould, Oliver and Hirvonen, Joonas",
    title = "{BubbleDet: a Python package to compute functional determinants for bubble nucleation}",
    eprint = "2308.15652",
    archivePrefix = "arXiv",
    primaryClass = "hep-ph",
    doi = "10.1007/JHEP12(2023)056",
    journal = "JHEP",
    volume = "2023",
    number = "12",
    pages = "056",
    year = "2023"
}

@article{Gould:2024chm,
    author = "Gould, Oliver and Kormu, Anna and Weir, David J.",
    title = "{Nonperturbative test of nucleation calculations for strong phase transitions}",
    eprint = "2404.01876",
    archivePrefix = "arXiv",
    primaryClass = "hep-th",
    reportNumber = "HIP-2024-9/TH",
    doi = "10.1103/PhysRevD.111.L051901",
    journal = "Phys. Rev. D",
    volume = "111",
    number = "5",
    pages = "L051901",
    year = "2025"
}

@article{Yang:1952be,
    author = "Yang, C. N. and Lee, T. D.",
    title = "{Statistical theory of equations of state and phase transitions. 1. Theory of condensation}",
    doi = "10.1103/PhysRev.87.404",
    journal = "Phys. Rev.",
    volume = "87",
    pages = "404--409",
    year = "1952"
}

@article{Lee:1952ig,
    author = "Lee, T. D. and Yang, C. N.",
    title = "{Statistical theory of equations of state and phase transitions. 2. Lattice gas and Ising model}",
    doi = "10.1103/PhysRev.87.410",
    journal = "Phys. Rev.",
    volume = "87",
    pages = "410--419",
    year = "1952"
}

@article{Mckane:1978me,
    author = "McKane, A. J.",
    title = "{Vacuum Instability in Scalar Field Theories}",
    doi = "10.1016/0550-3213(79)90086-5",
    journal = "Nucl. Phys. B",
    volume = "152",
    pages = "166--188",
    year = "1979"
}

@article{Jentschura:2010zza,
    author = "Jentschura, Ulrich D. and Surzhykov, Andrey and Zinn-Justin, Jean",
    title = "{Multi-instantons and exact results. III: Unification of even and odd anharmonic oscillators}",
    eprint = "1001.3910",
    archivePrefix = "arXiv",
    primaryClass = "math-ph",
    doi = "10.1016/j.aop.2010.01.002",
    journal = "Annals Phys.",
    volume = "325",
    pages = "1135--1172",
    year = "2010"
}

@article{Salvatori:2025oib,
    author = "Salvatori, Giulio",
    title = "{All loop scattering as a sampling problem}",
    eprint = "2503.07707",
    archivePrefix = "arXiv",
    primaryClass = "hep-th",
    doi = "10.1007/JHEP02(2026)196",
    journal = "JHEP",
    volume = "2026",
    number = "02",
    pages = "196",
    year = "2026"
}

@article{Giorgini:2021esl,
    author = "Giorgini, Ludovico T. and Jentschura, Ulrich D. and Malatesta, Enrico M. and Parisi, Giorgio and Rizzo, Tommaso and Zinn-Justin, Jean",
    title = "{Correlation functions of the anharmonic oscillator: Numerical verification of two-loop corrections to the large-order behavior}",
    eprint = "2111.12765",
    archivePrefix = "arXiv",
    primaryClass = "hep-th",
    doi = "10.1103/PhysRevD.105.105012",
    journal = "Phys. Rev. D",
    volume = "105",
    number = "10",
    pages = "105012",
    year = "2022"
}

@misc{Borinsky:2026sof,
    author = "Borinsky, Michael and Radkovski, Jury and Kannagi, Suriyah Rajalingam",
    title = "{Tropical QFT and the sine-Gordon model}",
    eprint = "2610.12653",
    archivePrefix = "arXiv",
    primaryClass = "hep-th",
    month = "10",
    year = "2026"
}

    \widetext
	\begin{center}
		\textbf{\large Supplemental Materials}
	\end{center}
	\appendix
	\setcounter{secnumdepth}{2}

\section{Tropical methods}
\label{app:tropical}

With the measure $\mathrm{d}^3k/\pi^{3/2}$ per loop, an $L$-loop graph $G$ with $E=3L-1$ internal edges contributes $y^L\Sigma_G(p)$ to the self-energy in Eq.~\eqref{eq:gamma2}, with $\Sigma_G$ given by its Schwinger representation in terms of the Kirchhoff polynomial $U_G(\alpha)$ and the external resistance $R_G(\alpha)$. Set $\alpha_e=\Lambda x_e$ with $\sum_ex_e=1$ and define $\hat R=R_G(x)=R_G(\alpha)/\Lambda$. The $\Lambda$ integral of the $p^{2k}$ term gives $\Gamma(\omega_L+k)$ with $\omega_L=E-3L/2=3L/2-1$, so that $d_L/s_L=-\omega_L\langle\hat R\rangle$ and $u_L/s_L=\frac12\omega_L(\omega_L+1)\langle\hat R^2\rangle$, averaged over the normalized projective integrand of all $L$-loop graphs; equivalently $d_L/s_L=-\langle R_G\rangle$ and $u_L/s_L=\frac12\langle R_G^2\rangle$ in the original variables. The vertex coefficients of $\Gamma^{(3)}(0)/g=1+\sum_Lt_Ly^L$ obey $t_L=\omega_Ls_L$, since $\Gamma^{(3)}(0)=g\,\partial[m_0^2-\Sigma(0)]/\partial m_0^2$.

\begin{figure}[t]
  \centering
  \includegraphics{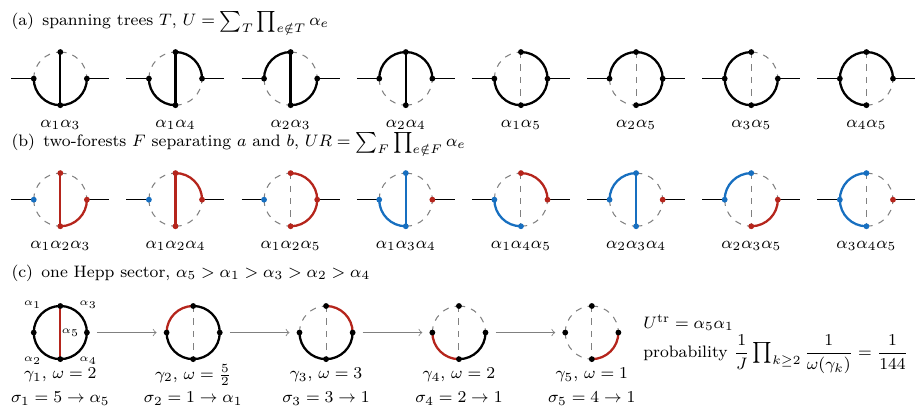}
  \caption{\label{fig:A:tropical}Tropical structures of the kite, with the labels of Fig.~\ref{fig:graphs}(d). (a)~The eight spanning trees (thick) and the monomials of their omitted edges (dashed), which sum to $U$. (b)~The eight two-forests separating $a$ (blue) from $b$ (red), whose omitted edges sum to $UR$. (c)~Sampling the Hepp sector $\alpha_5>\alpha_1>\alpha_3>\alpha_2>\alpha_4$. At each step the largest remaining parameter (red) is removed from $\gamma_k$. It contributes a factor $\alpha_{\sigma_k}$ to $U^{\rm tr}$ when it closes a loop of the smaller ones, and the parameters follow from $\alpha_{\sigma_k}=\alpha_{\sigma_{k-1}}\xi_k^{1/\omega(\gamma_k)}$.}
\end{figure}

Diagram tropical sampling~\cite{Borinsky:2020rqs} integrates one graph at a time. Consider a graph $G$ with $E$ edges, $L$ loops and external vertices $a,b$, with $U_G$ and $U_GR_G$ the sums over
spanning trees and over two-forests separating $a$ from $b$ given in the main text, listed for the kite in Fig.~\ref{fig:A:tropical}(a,b). For an edge subset $\gamma$ let
$L(\gamma)=|\gamma|-V(\gamma)+C(\gamma)$ be its loop number, where $V(\gamma)$ and $C(\gamma)$ count the vertices and
connected components of the subgraph formed by $\gamma$, and let $\omega(\gamma)=|\gamma|-\frac32L(\gamma)$ be its
superficial degree, with $\omega\equiv\omega(G)=E-\frac32L$. At $p=0$ the projective integrand is
$f=U_G^{-3/2}(\sum_e\alpha_e)^{-\omega}$ and $s_G=\Gamma(\omega)\,|\mathrm{Aut}\,G|^{-1}\int\Omega\,f$. Here $\Omega$ denotes the projective measure, represented on $\sum_e\alpha_e=1$ by $\delta(1-\sum_e\alpha_e)\prod_e\mathrm{d}\alpha_e$. In the Hepp sector
$\alpha_{\sigma_1}>\alpha_{\sigma_2}>\dots>\alpha_{\sigma_E}$ let $\gamma_k=\{\sigma_k,\dots,\sigma_E\}$, with
$\gamma_{E+1}=\emptyset$. The largest monomial of $U_G$ is then
\begin{equation}\label{eq:A:Utr}
  U^{\rm tr}_G=\prod_{k=1}^{E}\alpha_{\sigma_k}^{\,L(\gamma_k)-L(\gamma_{k+1})},\qquad
  U^{\rm tr}_G\le U_G\le N_T(G)\,U^{\rm tr}_G,
\end{equation}
so that $\sigma_k$ is left out of the spanning tree exactly when it closes a loop among the smaller parameters, and $N_T$
is the number of spanning trees~\cite{Borinsky:2020rqs}. Over the sector the tropical integrand
$f^{\rm tr}=(U^{\rm tr}_G)^{-3/2}\alpha_{\sigma_1}^{-\omega}$ integrates to $\prod_{k=2}^{E}1/\omega(\gamma_k)$. Summed over
all sectors this gives $J(G)$, which follows from the recursion
\begin{equation}\label{eq:A:J}
  J(\gamma)=\sum_{e\in\gamma}\frac{J(\gamma\setminus e)}{\omega(\gamma\setminus e)},\qquad J(\{e\})=1,
\end{equation}
over the $2^E$ edge subsets. A sample is generated in three steps. (i) Starting from the full edge set $\gamma_1$, for
$k=1,\dots,E-1$ remove $\sigma_k=e$ from $\gamma_k$ with probability $J(\gamma_k\setminus e)/[\omega(\gamma_k\setminus
e)\,J(\gamma_k)]$ and set $\gamma_{k+1}=\gamma_k\setminus\sigma_k$; the remaining edge is $\sigma_E$. This selects each
sector with probability $J(G)^{-1}\prod_{k\ge2}1/\omega(\gamma_k)$. (ii) Set $\alpha_{\sigma_1}=1$ and
$\alpha_{\sigma_k}=\alpha_{\sigma_{k-1}}\,\xi_k^{1/\omega(\gamma_k)}$ for $k=2,\dots,E$, with independent uniform
$\xi_k\in(0,1)$. This samples $f^{\rm tr}$ within the sector exactly. Figure~\ref{fig:A:tropical}(c) follows these steps for one sector of the kite. (iii) Evaluate $U_G$, $R_G$ and
Eq.~\eqref{eq:A:Utr} at these $\alpha$ and form the weight $w=(U^{\rm tr}_G/U_G)^{3/2}(\sum_e\alpha_e)^{-\omega}\in(0,1]$.
Since $R_G$ is homogeneous of degree one, the $p^2$ and $p^4$ terms carry $\Gamma(\omega+1)$ and $\Gamma(\omega+2)$, and
averaging over the samples gives
\begin{equation}\label{eq:A:diag}
  s_G=\frac{\Gamma(\omega)\,J(G)}{|\mathrm{Aut}\,G|}\,\mathbb{E}[w],\qquad
  d_G=-\frac{\Gamma(\omega+1)\,J(G)}{|\mathrm{Aut}\,G|}\,\mathbb{E}\Bigl[\frac{w\,R_G}{\sum_e\alpha_e}\Bigr],\qquad
  u_G=\frac{\Gamma(\omega+2)\,J(G)}{2\,|\mathrm{Aut}\,G|}\,\mathbb{E}\Bigl[\frac{w\,R_G^2}{(\sum_e\alpha_e)^2}\Bigr].
\end{equation}
At two loops $J=48/5$ for the kite and $76/5$ for the self-energy insertion, and with $10^5$ samples per graph the estimators reproduce the exact two-loop values within the statistical errors. The kite, for example, gives $s=0.1746(6)$, $d=-0.0517(2)$ and $u=0.01302(4)$, against $\pi/18$, $-4\pi/243$ and $1451\pi/349920$.

\begin{figure}[t]
  \centering
  \includegraphics{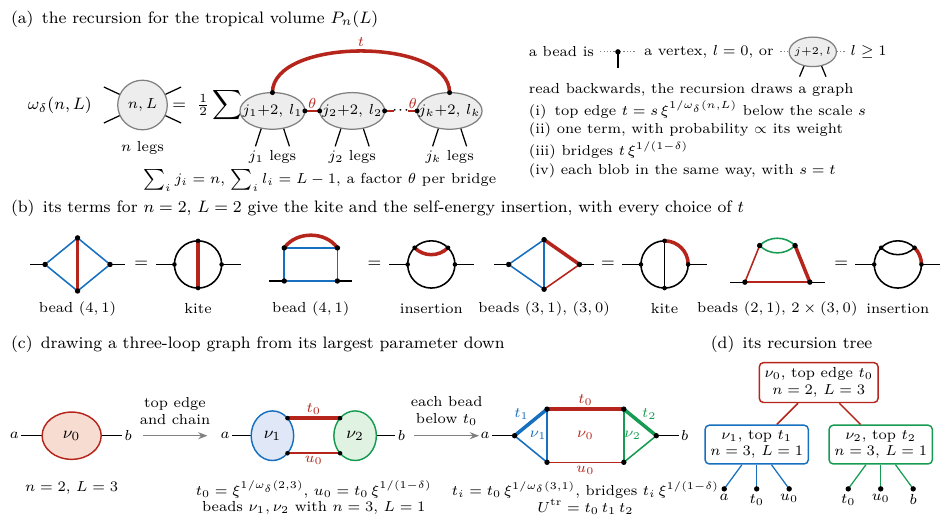}
  \caption{\label{fig:A:global}Global tropical sampling. (a)~The recursion~\eqref{eq:A:rec} as a picture. A blob stands for all 1PI graphs with $n$ legs and $L$ loops, weighted by $\mu_\delta$. The largest parameter $t$ (red) closes a cycle through a chain of beads joined by bridges. Each bead is a vertex or a smaller blob below $t$. (b)~Terms for $n=2$ and $L=2$, with the beads opened. A single bead $(4,1)$ gives the kite or the self-energy insertion, depending on the order of its legs. Together the terms give $P_2(2)$ as the sum of the two graphs, each with every choice of its largest edge. (c)~A draw of a three-loop graph. The top edge $t_0$ closes a chain of the triangles $\nu_1$ and $\nu_2$ joined by the bridge $u_0$, and each triangle is drawn in the same way below $t_0$. Colors mark the cycle of each node, and its top edge is thick. (d)~The recursion tree of this draw. Its leaves are vertices, labelled by the leg they carry, and $U^{\rm tr}$ is the product of the top edges.}
\end{figure}

\begin{figure}[t]
  \centering
  \includegraphics{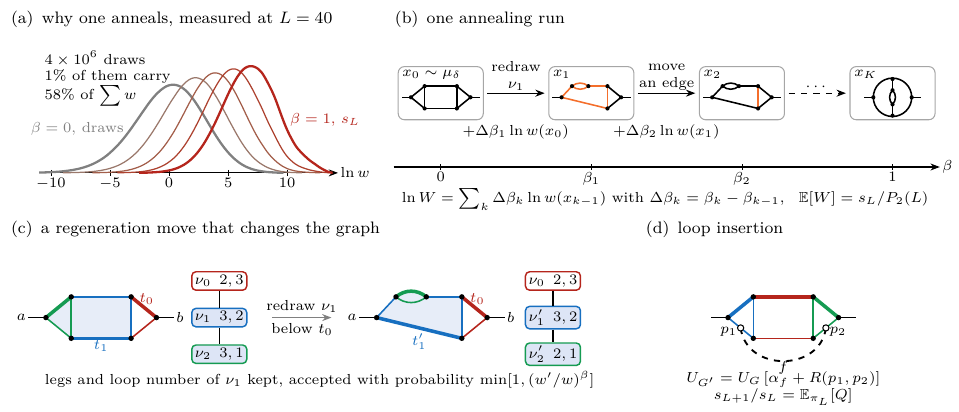}
  \caption{\label{fig:A:anneal}Annealed sampling. (a)~Distribution of $\ln w$, relative to its mean, over $4\times10^6$ direct draws at $L=40$, weighted by $w^\beta$ for $\beta=0,\frac14,\frac12,\frac34,1$. The draws follow $\beta=0$, while $s_L=P_2(L)\,\mathbb{E}[w]$ comes from $\beta=1$, where few draws fall. One percent of the draws carry 58\% of $\sum w$. (b)~One annealing run. Each frame is one state, a graph together with all its parameters. Each step adds $\Delta\beta_k\ln w$ to $\ln W$ and then moves the state with a Markov kernel that leaves $\pi_{\beta_k}$ invariant. Changed parts are orange. (c)~The first move of the run. The subtree of $\nu_1$ (shaded) is redrawn below $t_0$, and its triangle becomes a bubble. The parameters are ordered differently from Fig.~\ref{fig:A:global}(c), so the same graph has a different tree. (d)~Inserting an edge $f$ between points $p_1$ and $p_2$ adds a loop and gives the ratio $s_{L+1}/s_L$.}
\end{figure}

Global tropical sampling~\cite{Borinsky:2025stz} instead draws graphs and parameters jointly. In cube variables $u_e=1-e^{-\alpha_e}$, with $e^{-\alpha}\mathrm{d}\alpha=\mathrm{d}u$, we use the tilted measure
\begin{equation}\label{eq:A:mu}
  \mu_\delta(G,\mathrm{d}u)=\frac{1}{|\mathrm{Aut}\,G|}\prod_eu_e^{-\delta}\;U^{\rm tr}_G(u)^{-3/2}\prod_e\mathrm{d}u_e,
  \qquad P_n(L)=\sum_G\int\mu_\delta ,
\end{equation}
summed over 1PI graphs with $n$ labelled legs and $L$ loops. Removing the edge with the largest parameter $t$ leaves a chain of 1PI beads joined by bridges (Fig.~\ref{fig:A:global}(a)). That edge closes a cycle and contributes a factor $t$ to $U^{\rm tr}$, $U^{\rm tr}$ factorizes over the beads, bridges lie in every spanning tree, and the measure is homogeneous in $t$ of degree $\omega_\delta(n,L)=(1-\delta)(3L-3+n)-3L/2$. Hence
\begin{equation}\label{eq:A:rec}
  \omega_\delta(n,L)\,P_n(L)=\tfrac12\,\mathrm{ch}(n,L-1),\qquad
  \mathrm{ch}(n,l)=n!\,[x^nz^l]\,\frac{B}{1-\theta B},\qquad
  B(x,z)=\sum_{j,l}P_{2+j}(l)\,\frac{x^jz^l}{j!},
\end{equation}
with bridge factor $\theta=1/(1-\delta)$, $P_3(0)=1$ for a single vertex with one external leg, and $\frac12$ for the orientation of the chain. Positivity of $\omega_\delta(2,1)=\frac12-2\delta$ requires $\delta<1/4$. At $\delta=0$ this is Borinsky's tropical loop equation~\cite{Borinsky:2025stz} order by order, equivalent to the tropical Feynman rules of Ref.~\cite{Balduf:2025idm}. Figure~\ref{fig:A:global}(a) shows it as a picture, and Fig.~\ref{fig:A:global}(b) shows how its terms at $n=2$ and $L=2$ build the two two-loop graphs. Its tables cost $O(L^4)$ operations, and read in reverse they give an exact sampler of graphs, sectors and parameters. A node with $n$ legs and $L$ loops below the scale $s$, with $s=1$ at the root, draws its top edge $t=s\,\xi^{1/\omega_\delta(n,L)}$ with a fresh uniform $\xi\in(0,1)$. It picks a chain with probability proportional to its term in $\mathrm{ch}(n,L-1)$ and draws each bridge as $t\,\xi^{1/(1-\delta)}$. A bead with $l\ge1$ loops that carries $j$ of the $n$ legs is a node with $j+2$ legs and $l$ loops below $t$, and a bead without loops is a vertex that carries one leg (Fig.~\ref{fig:A:global}(c)). The nodes form a recursion tree, and $U^{\rm tr}$ is the product of their top edges (Fig.~\ref{fig:A:global}(d)). A Kron reduction of the weighted Laplacian then gives $U(\alpha)$ and the external resistance $R_{12}$, and
\begin{equation}\label{eq:A:est}
  w=\Bigl[\frac{U^{\rm tr}(u)}{U(\alpha)}\Bigr]^{3/2}\prod_eu_e^{\delta}\in(0,1],\qquad
  s_L=P_2(L)\,\mathbb{E}[w],\quad d_L=-P_2(L)\,\mathbb{E}[wR_{12}],\quad u_L=\tfrac12P_2(L)\,\mathbb{E}[wR_{12}^2].
\end{equation}
At $3\le L\le18$ the three-point weight has relative variance 0.5--23, against 15--$3\times10^9$ implied by the errors quoted for the projective estimator of Ref.~\cite{Borinsky:2025stz}. However, $\mathrm{Var}(\ln w)=0.181L+1.33$, the relative variance grows by a factor 1.2 per loop, and for $L\gtrsim25$ the Hill tail index of $w$ falls below 2. A few rare draws then carry most of $\sum w$ (Fig.~\ref{fig:A:anneal}(a)), so direct sampling stalls near $L\approx50$--60.

Beyond this range we use annealed importance sampling~\cite{Neal:2001hwj,Jarzynski:1996oqb}, which acts on the recursion tree and its parameters (Fig.~\ref{fig:A:anneal}(b)). Along
\begin{equation}\label{eq:A:path}
  \pi_\beta\propto\mu_\delta\,w^\beta,\quad \beta_k=(k/K)^p,\qquad
  \mathbb{E}[W]=\frac{s_L}{P_2(L)},\quad \mathbb{E}[W\,f(x_K)]=\frac{s_L}{P_2(L)}\,\mathbb{E}_{\pi_1}[f],
\end{equation}
with $K=L$, $p=1$ or 0.75 and $\delta=0$, a run starts from an exact tropical draw and at each step adds $(\beta_k-\beta_{k-1})\ln w$ to $\ln W$ and applies a Markov kernel that leaves $\pi_{\beta_k}$ invariant. Edge moves redraw one parameter, and regeneration moves redraw the subtree of a random node below the top edge of its parent, with its legs and loop number fixed (Fig.~\ref{fig:A:anneal}(c)). Both draw from the exact tropical conditional, so the Metropolis ratio is $(w'/w)^\beta$. With cached Schur complements of the subgraph Laplacians a sweep costs $\propto L^{3.5\text{--}3.8}$. By Neal's extended-space argument the identities in Eq.~\eqref{eq:A:path} hold for any schedule and degree of mixing, with $f=-R_{12}$ for $d_L$. Errors from independent runs are reliable for $\mathrm{Var}\ln W\lesssim0.6$, where the Hill index of $W$ exceeds 2. Each run $r$ ends with 96 sweeps at $\beta=1$. These sweeps leave $\pi_1$ invariant, so Eq.~\eqref{eq:A:path} holds for each of them. The insertion observable $Q$, the bracket in Eq.~\eqref{eq:A:lir} below, is averaged over them to $\bar Q_r$. The runs are pooled with their weights, $s_{L+1}/s_L=\sum_rW_r\bar Q_r/\sum_rW_r$, with delta-method errors. The ratio $d_{L+1}/s_L$ is estimated in the same way.

Ratios of consecutive orders follow from a loop-insertion identity. Inserting an edge $f$ between points $p_1$ and $p_2$, at positions $\tau_1$ and $\tau_2$ along edges $e_1$ and $e_2$ of a sampled $L$-loop graph (Fig.~\ref{fig:A:anneal}(d)), gives $U_{G'}=U_G\,[\alpha_f+R(p_1,p_2)]$, and $\alpha_f$ integrates to $J(R)=\int_0^\infty\mathrm{d}\alpha\,e^{-\alpha}(\alpha+R)^{-3/2}=2/\sqrt R-2\sqrt\pi\,e^R\,\mathrm{erfc}\sqrt R$. Every labelled $(L{+}1)$-loop graph $G'$ with an oriented removable edge, an internal edge without external endpoint whose removal leaves $G'$ bridgeless, arises from exactly one insertion, so
\begin{equation}\label{eq:A:lir}
  \frac{s_{L+1}}{s_L}=\mathbb{E}_{\pi_L}\Biggl[\frac12\sum_{(e_1,e_2)}\int_0^{\alpha_{e_1}}\!\!\mathrm{d}\tau_1\int_0^{\alpha_{e_2}}\!\!\mathrm{d}\tau_2\;\frac{J(R(p_1,p_2))}{m(G')}\Biggr],
\end{equation}
with $\pi_L$ the normalized $L$-loop integrand and $m(G')\ge1$ the number of removable edges, computed exactly from the two-edge-cut classes of $G$. The sum runs over ordered pairs of edges, including $e_1=e_2$, so $p_1$ and $p_2$ each range over the whole graph, and the factor $\frac12$ removes their exchange. The Sherman--Morrison update $R_{ab}-\Delta^2/(\alpha_f+R)$ of the external resistance gives $d_{L+1}/s_L$ in the same way, with $J$ replaced by $-[R_{ab}J(R)-\Delta^2J_{5/2}(R)]$, where $\Delta$ is the potential difference between $p_1$ and $p_2$ and $J_{5/2}$ has the power $-5/2$. Consecutive ratios thus cost $10^3$--$4\times10^3$ times less than ratios of independent absolute values. Absolute coefficients follow by telescoping, $s_{L+1}=s_2\prod_{l=2}^{L}s_{l+1}/s_l$ from the exact $s_2=\pi/6$, with full covariance.

Per unit relative precision the cost of a consecutive ratio grows as $L^{4.7}$. All orders through $L=325$ required about $1.6\times10^5$ CPU core-hours in total. The methods reproduce the exact $s_1$, $d_1$, $s_2$, $d_2$ and $s_2/s_1$ to $10^{-5}$.

\section{The instanton through four loops}
\label{app:instanton}

This appendix computes the instanton side of Table~\ref{tab:inst}, the one-loop factor $C_s$ and the fluctuation corrections $\hat\kappa_1=-14.36586(3)$, $\hat\kappa_2=45.4506(5)$ and $\hat\kappa_3=-48.15(40)$ of Eq.~\eqref{eq:loops}, where $\hat\kappa_k$ comes from $k+1$ loops around the instanton and $S_{\rm cl}=S_0/g^2=1/(ay)$. We first relate them to the coefficients, then construct the instanton and its one-loop factor, and finally the Feynman rules and graphs that give the higher loops.

The large orders follow from a dispersion relation. For $y>0$ every coefficient function $F(y)=\sum_LF_Ly^L$ of the self-energy of Eq.~\eqref{eq:gamma2} in powers of $p^2$ has a cut along the positive axis, and $F_L=\pi^{-1}\int_0^\infty\mathrm{d}y\,y^{-L-1}\,\mathrm{Im}\,F(y+i0)$. If $\mathrm{Im}\,F=\hat C\,y^{-b}e^{-1/(ay)}[1+\sum_k\hat\kappa_k(ay)^k]$, term-by-term integration gives $F_L=\pi^{-1}\hat C\,a^{L+b}[\Gamma(L+b)+\sum_k\hat\kappa_k\Gamma(L+b-k)]$, which for $s_L$, with $b=5/2$, is
\begin{equation}\label{eq:B:RL}
  s_L=C_s\,a^L\,\Gamma\bigl(L+\tfrac52\bigr)R_L,\qquad
  R_L=1+\sum_{k\ge1}\frac{\hat\kappa_k}{(L+\frac32)(L+\frac12)\cdots(L+\frac52-k)} ,\qquad C_s=\frac{\hat C a^{5/2}}{\pi}.
\end{equation}
Each loop thus brings one inverse power of $L$. The three translation zero modes give $S_{\rm cl}^{3/2}\propto y^{-3/2}$ and the two zero-momentum insertions of the classical field a further $y^{-1}$, hence $b=5/2$. The power basis $R_L=1+\sum_kc_kL^{-k}$ mixes loop orders,
\begin{equation}\label{eq:B:basis}
  \hat\kappa_1=c_1,\quad \hat\kappa_2=c_2+\tfrac32c_1,\quad \hat\kappa_3=c_3+2c_2+\tfrac34c_1,\quad
  \hat\kappa_4=c_4+\tfrac32c_3-\tfrac14c_2-\tfrac38c_1 .
\end{equation}

The instanton is $\phi_{\rm cl}=\psi(r)/g$, with $m_0=1$ and
\begin{equation}\label{eq:B:bounce}
  \psi''+\frac2r\psi'=\psi+\frac12\psi^2,\qquad\psi'(0)=0,\qquad\psi(r)\to A_\psi\frac{e^{-r}}{r}\quad(r\to\infty).
\end{equation}
Shooting gives $\psi(0)=-8.383365908885$, $\psi''(0)=8.919015357$, $A_\psi=-32.13885535749$ and $S_0=174.640946865$, with three evaluations of the action agreeing to $10^{-13}$. The field equation and stationarity under dilations give $\int\psi^2=\int(\nabla\psi)^2=3S_0$, $\int\psi^3=-12\,S_0$ and $\int\psi=-\frac32S_0$, which hold numerically to $10^{-13}$. The only complex radial saddles we find are one conjugate pair with $S/S_0=8.42\pm8.47i$, suppressed in the coefficients by $11.9^{-L}$.

The fluctuations $\eta$, with $\phi=\phi_{\rm cl}(x-x_0)+\eta$ and $x_0$ the position of the instanton, have an exactly cubic action,
\begin{equation}\label{eq:B:fluct}
  S[\phi]=S_{\rm cl}+\tfrac12\langle\eta,M\eta\rangle+\frac g6\!\int\!\eta^3+h\!\int\!(\phi_{\rm cl}+\eta),\qquad M=-\nabla^2+1+\psi(r).
\end{equation}
The operator $M$ has one negative mode at $l=0$, with eigenvalue $-2.364$, three zero modes $\partial_\mu\phi_{\rm cl}$ at $l=1$, and a continuum from 1. The negative mode produces the imaginary part, and the zero modes are traded for $x_0$. At one loop
\begin{equation}\label{eq:B:oneloop}
  \mathrm{Im}\,\Sigma(p;y)=\frac{1}{2}\Bigl(\frac{S_{\rm cl}}{2\pi}\Bigr)^{3/2}\frac{e^{-S_{\rm cl}}}{|\det_2'|^{1/2}}\,(1+p^2)^2\,\frac{\tilde\psi(p)^2}{g^2},
\end{equation}
with the half-Gaussian $\frac12$ along the negative mode~\cite{Callan:1977pt}, the Jacobian $(S_{\rm cl}/2\pi)^{3/2}$ of the zero modes, the amputation $(1+p^2)^2$, and the Carleman determinant $\det'_2(M/M_0)$ without zero modes, $M_0=-\nabla^2+1$, which remains after the counterterm cancels the tadpole $\mathrm{Tr}\,G_0\psi$, $G_0=M_0^{-1}$. In partial waves~\cite{Gelfand:1959nq,Dunne:2006ct,Dunne:2005rt}, with $T_l(\infty)=\det(M_l/M_l^0)$ the Gel'fand--Yaglom ratio of the radial operators,
\begin{equation}\label{eq:B:GY}
  \ln|\det'_2|=\sum_l(2l+1)\bigl[\ln|T_l(\infty)|-b_l-c_l\bigr]+C_2 ,
\end{equation}
where $b_l$ and $c_l$ are the first two Born terms and $C_2=-\frac12\mathrm{Tr}(G_0\psi G_0\psi)=-9.73407178$. At $l=0$, $T_0(\infty)=-0.0636654$ reflects the negative mode, and at $l=1$ the zero modes are removed in closed form~\cite{Dunne:2005rt}, $\det'(M_1)/\det(M_1^0)=S_0/[4\pi\psi''(0)|A_\psi|]$. Three evaluations agree on $\ln|\det'_2|=-6.13045499(1)$. With $\tilde\phi_{\rm cl}(0)=-\frac32S_0/g$ and Eq.~\eqref{eq:B:RL},
\begin{equation}\label{eq:B:Cs}
  C_s=\frac{9S_0}{8\pi(2\pi)^{3/2}|\det{}'_2|^{1/2}}=85.1315052 .
\end{equation}
Define the large-order limits $\rho_d=\lim_{L\to\infty}d_L/s_L$ and $\rho_4=\lim_{L\to\infty}u_L/s_L$. At nonzero momentum, $(1+p^2)\tilde\psi=-\frac12\widetilde{\psi^2}$, and the moments $\mu_{2n}=\int r^{2n}\psi/\!\int\psi$ give
\begin{equation}\label{eq:B:rho}
  \rho_d=2-\frac{\mu_2}{3},\qquad
  \rho_4=1-\frac{2\mu_2}{3}+\frac{\mu_2^2}{36}+\frac{\mu_4}{60},
\end{equation}
so that $\mu_2=8.4350624$ and $\mu_4=181.39608$ give $\rho_d=-0.8116875$ and $\rho_4=0.3762897$. The prediction for $\rho_4$ was made before the $p^4$ data were analysed.

Beyond one loop, $\hat\kappa_k$ follows from connected vacuum graphs of $\eta$. We write the discontinuity of the vacuum free energy, normalized by its one-loop value, as
\begin{equation}\label{eq:B:ell}
  1+\sum_{n\ge1}\frac{v_n}{S_{\rm cl}^n}=\exp\Bigl(\sum_{n\ge1}\frac{\ell_n}{S_{\rm cl}^n}\Bigr),
\end{equation}
where $\ell_n$ is $S_0^n$ times the connected vacuum graphs of order $g^{2n}$, the $(n{+}1)$-loop graphs, minus the same graphs around the perturbative vacuum. Since $g^2=S_0/S_{\rm cl}$, each loop costs one power of $1/S_{\rm cl}$. The two-point function reduces to these by an exact chain rule. In $\phi^3$ theory a constant source $h$ is absorbed by shifting $\phi$ to the minimum of $U(\phi)=\frac12\phi^2+\frac g6\phi^3+h\phi$, which leaves the same theory with mass $m^2=\sqrt{1-2gh}$. The vacuum discontinuity is therefore $D(m^2)=m^3\hat D(y\,m^{-3})$, and its derivatives with respect to the source, followed by the Legendre transform at $\langle\phi\rangle=0$, give the self-energy with the external fields, the counterterm shift and the amputation included exactly,
\begin{equation}\label{eq:B:chain}
  \hat\kappa_1=v_1-\tfrac{16}{3},\qquad \hat\kappa_2=v_2-\tfrac{10}{3}v_1+\tfrac{55}{12},\qquad
  \hat\kappa_3=v_3-\tfrac43v_2+\tfrac54v_1 .
\end{equation}
Numerically $c_2=\ell_2+97.034$ and $c_3=\ell_3-0.396$. The chain rule holds exactly in zero dimensions and reproduces exact one-dimensional series to $10^{-6}$--$10^{-17}$.

\begin{figure}[!htbp]
  \centering
  \includegraphics{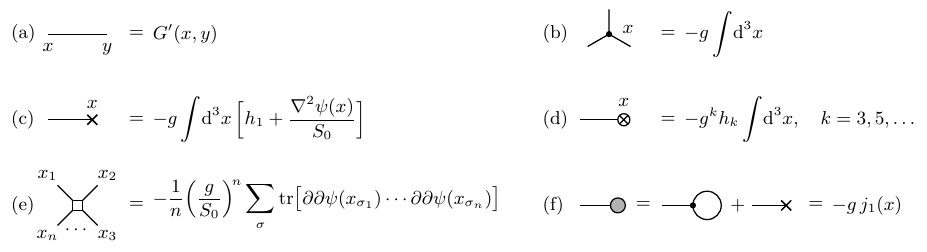}
  \caption{\label{fig:B:rules}Feynman rules for the fluctuations around the instanton, in the spectral gauge. (a)~The propagator. (b)~The cubic vertex. (c)~The one-leg vertex of order $g$, the counterterm $h_1$ together with $\mathrm{tr}X$. (d)~The higher counterterms. (e)~The Faddeev--Popov vertex $\mathrm{tr}X^n/n$, whose legs end at points $x_i$ that are integrated over, with the indices contracted around the vertex. (f)~The cubic tadpole and the order-$g$ vertex add up to the finite one-point vertex.}
\end{figure}

The Feynman rules are collected in Fig.~\ref{fig:B:rules}. The position $x_0$ is fixed by the Faddeev--Popov condition $\langle\partial_\mu\Phi(\cdot-x_0),\phi\rangle=0$ for a radial gauge function $\Phi$~\cite{Zinn-Justin:2002ecy}. In the spectral gauge $\Phi=\phi_{\rm cl}$ the measure of $x_0$ is $(S_{\rm cl}/2\pi)^{3/2}\det(1-X)\,\mathrm{d}^3x_0$, with $X_{\mu\nu}=(g/S_0)\langle\partial_\mu\partial_\nu\psi,\eta\rangle$, and
\begin{equation}\label{eq:B:FP}
  \det(1-X)=\exp\Bigl(-\sum_{n\ge1}\frac1n\,\mathrm{tr}X^n\Bigr)
\end{equation}
adds Faddeev--Popov vertices $\mathrm{tr}X^n/n$ with $n$ legs. The propagator $G'$ is the inverse of $M$ on the fluctuations that satisfy the gauge condition, built from the radial solutions of Eq.~\eqref{eq:B:GY} in partial waves. At $l=0$ the negative eigenvalue enters with its sign, as Gaussian moments along the steepest-descent contour require, and at $l=1$ the zero modes are projected out. In the spectral gauge $M(x_\mu\partial_\nu\psi)=-2\,\partial_\mu\partial_\nu\psi$, so that
\begin{equation}\label{eq:B:Pid}
  G'\,\partial_\mu\partial_\nu\psi=-\tfrac12\,x_\mu\partial_\nu\psi,\qquad
  \langle\partial_\alpha\partial_\beta\psi,G'\,\partial_\gamma\partial_\delta\psi\rangle=\frac{S_0}{4}\bigl(\delta_{\alpha\gamma}\delta_{\beta\delta}+\delta_{\alpha\delta}\delta_{\beta\gamma}-\delta_{\alpha\beta}\delta_{\gamma\delta}\bigr),
\end{equation}
and every Faddeev--Popov leg contracts in closed form. The counterterm is fixed by the Schwinger--Dyson equation, $h=-\frac g2\langle\phi^2\rangle=gh_1+g^3h_3+\cdots$ in the perturbative vacuum, with $h_1=-\frac12G_0(0)$. Its one-leg vertices combine with the cubic tadpole into the finite one-point vertex $-g\,j_1$, with
\begin{equation}\label{eq:B:j1}
  j_1(x)=\tfrac12\Delta G(x)+\frac{\nabla^2\psi(x)}{S_0},\qquad \Delta G(x)=G'(x,x)-G_0(x,x),
\end{equation}
where the second term is the one-leg vertex $\mathrm{tr}X$, and its value on the background, $h\int\phi_{\rm cl}$, cancels exactly the part of the 1PI cubic graphs linear in $\psi$.

\begin{figure}[!htbp]
  \centering
  \includegraphics{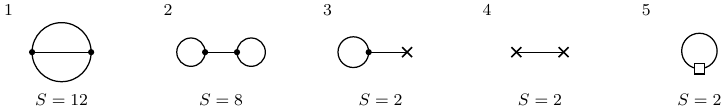}
  \caption{\label{fig:B:graphs}The five two-loop vacuum graphs, with their symmetry factors $S$. Graphs 2--4 combine into one-point vertices, Fig.~\ref{fig:B:rules}(f). The 24 three-loop and 136 four-loop graphs are built from the same elements.}
\end{figure}

At two loops the five graphs of Fig.~\ref{fig:B:graphs} give, in units of $1/S_{\rm cl}$ and in the spectral gauge, $v_1=V_{\rm sun}+V_{\rm dumb}+V_{J1}+V_{J2}$ with
\begin{equation}\label{eq:B:twoloop}
  V_{\rm sun}=\frac{S_0}{12}\int\!\!\int\bigl[G'^3-G_0^3-3G_0^2\,\delta G\bigr],\qquad
  V_{\rm dumb}=\frac{S_0}{8}\bigl\langle\Delta G,G'\Delta G\bigr\rangle,\qquad
  V_{J1}=-\frac14\bigl\langle x\cdot\nabla\psi,\Delta G\bigr\rangle,\qquad V_{J2}=-\frac32 .
\end{equation}
From the sunset, graph 1, we subtract its free value and its part linear in $\psi$, with $\delta G=-G_0\psi G_0$. Graphs 2--4 give $\frac12g^2\langle j_1,G'j_1\rangle$, whose terms quadratic and linear in $\Delta G$ are $V_{\rm dumb}$ and $V_{J1}$, and $V_{J2}$ collects the remaining term, $-\frac38$, and the Faddeev--Popov loop of graph 5, $-\frac98$, both exact by Eq.~\eqref{eq:B:Pid}. With $V_{\rm sun}=-3.01932(3)$, $V_{\rm dumb}=1.098062$ and $V_{J1}=-5.611269$, $v_1=-9.03253(3)$ and $\hat\kappa_1=-14.36586(3)$. 

At three and four loops the same rules generate 24 and 136 graphs, which regroup in the same way into 12 and 60 graphs to evaluate in three dimensions. We evaluate them by two independent methods. The first is deterministic. Each propagator is expanded in partial waves, the angular integrals reduce to 3j, 6j and 9j symbols, the radial integrals become nested one-dimensional integrals, and subgraphs made of $G_0$ alone are done in momentum space. The partial-wave sums are extrapolated. At four loops the two hardest graphs, summed to $l=11$, set the error. A geometric and a power-law continuation of their tails differ by 0.43, and we quote the midpoint with half the difference. The second method samples the positions of the vertices by Monte Carlo, with $G'-G_0$ from partial waves up to $l=40$ and the Faddeev--Popov legs from Eq.~\eqref{eq:B:Pid}. Its weights are heavy-tailed, so errors come from independent runs and are inflated by a factor 1.4. The methods give $\ell_2=-30.0344(2)$ and $-30.03(8)$ and $\ell_3=-170.95(39)$ and $-170.9(8)$, hence, with the two gauges combined at four loops, $c_2=66.9994(5)$ and $c_3=-171.37(40)$.

Individual graphs depend on the gauge, and only their sum at each order does not. Changing $\Phi$ changes the $l=1$ part of $G'$ and the Faddeev--Popov vertices, which tests both. At three loops $\ell_2$ agrees to $10^{-4}$ between $\Phi=\psi$, $\psi^2$ and $e^{-r^2/2}$, while single contributions move by up to 15. At four loops the 1PI graphs with cubic vertices only move by $+82.2$ from $\Phi=\psi$ to $\psi^2$ and all other graphs by $-82.3$, and $\ell_3$ changes only from $-170.95(39)$ to $-171.01(57)$. In zero dimensions, where the instanton is $\psi=-2$ with $S_0=\frac23$ and no zero modes, the graph sums give $\ell_1=-\frac79$, $\ell_2=-\frac{10}{27}$ and $\ell_3=-\frac{614}{243}$ exactly and reproduce the large-order behavior of the graph counts. For the cubic anharmonic oscillator~\cite{Jentschura:2010zza}, with $\psi=-3\,\mathrm{sech}^2(t/2)$ and one zero mode, both implementations reproduce the exact series through four loops to $10^{-12}$--$10^{-9}$.

Table~\ref{tab:B:results} summarizes the results. The slope coefficients $e_k$ of $d_L/s_L=\rho_d(1+\sum_ke_kL^{-k})$, with $e_1=\Delta c_1=c_1(d)-c_1(s)$ the difference of the power-basis coefficients of $d_L$ and $s_L$, need in addition the second moments of the one-point function and of the propagator in the instanton background. The coefficients $e_2$ and $e_3$ come from a single implementation. The semiclassical calculations contain no adjustable parameters, but they were not carried out blind to the diagrammatic data.

\begin{table}[t]
\caption{\label{tab:B:results}Semiclassical coefficients in the power basis $c_k$, the factorial basis $\hat\kappa_k$ of
Eq.~\eqref{eq:B:RL}, and $d_L/s_L=\rho_d(1+\sum_ke_kL^{-k})$. ``Impl.'' is the number of independent implementations.}
\begin{ruledtabular}
\begin{tabular}{llcl}
quantity & value & impl. & method\\
\hline
$c_1=\hat\kappa_1$ (2 loops) & $-14.36586(3)$ & 2 & exact pieces ($-41/6$) and three vacuum integrals\\
$e_1=\Delta c_1$ & $3.053872$ & 2 & insertion integrals, five gauges\\
$c_2$ (3 loops) & $66.9994(5)$ & 2 & partial waves, three gauges; Monte Carlo $67.00(9)$\\
$c_3$ (4 loops) & $-171.37(40)$ & 2 & partial waves, two gauges; Monte Carlo $-171.4(1.6)$\\
$e_2$, $e_3$ & $8.72(15)$, $-5.37(3.63)$ & 1 & Monte Carlo\\
$\hat\kappa_2$, $\hat\kappa_3$ & $45.4506(5)$, $-48.15(40)$ & & Eq.~\eqref{eq:B:basis}\\
\end{tabular}
\end{ruledtabular}
\end{table}

\begin{table}[!htbp]
\caption{\label{tab:C:coeffs}Coefficients of the self-energy $\Sigma(p)=\sum_Ly^L(s_L+d_Lp^2+u_Lp^4+\cdots)$. $L=1,2$ are
exact, $3\le L\le325$ telescoped, with relative $1\sigma$ errors.
$R_L=s_L/[C_sa^L\Gamma(L+\frac52)]$ uses the instanton values $a=0.2550754836$ and $C_s=85.1315052$, with no fit.
The ratio $u_L/s_L$ is listed for $L\le2$, where it is exact, and for $L=20$--100, where it was measured.
The third column gives $s_L=t_L/(\frac32L-1)$ from the three-point coefficients $t_L$ of Ref.~\cite{Borinsky:2025stz}.}
\centering\footnotesize
\begin{tabular}{@{}r@{\quad}l@{\,}l@{\quad}l@{\,}l@{\quad}l@{\,}l@{\quad}l@{\,}l@{\quad}l@{\,}l@{\quad}l@{\quad}l@{}}
\toprule
$L$ & \multicolumn{2}{l}{$s_L$} & \multicolumn{2}{l}{$s_L$, Ref.~\cite{Borinsky:2025stz}} & \multicolumn{2}{l}{$\delta s_L/s_L$} & \multicolumn{2}{l}{$d_L$} & \multicolumn{2}{l}{$\delta d_L/|d_L|$} & $u_L/s_L$ & $R_L$\\
\midrule
1 & $8.8622693$ & $\times10^{-1}$ & $8.862218(22)$ & $\times10^{-1}$ & \multicolumn{2}{l}{exact} & $-7.3852244$ & $\times10^{-2}$ & \multicolumn{2}{l}{exact} & $1/80$ & $0.012280$\\
2 & $5.2359878$ & $\times10^{-1}$ & $5.235955(30)$ & $\times10^{-1}$ & \multicolumn{2}{l}{exact} & $-1.4706015$ & $\times10^{-1}$ & \multicolumn{2}{l}{exact} & $10537/145800$ & $0.008127$\\
3 & $8.2919257$ & $\times10^{-1}$ & $8.29197(10)$ & $\times10^{-1}$ & $9.1$ & $\times10^{-6}$ & $-3.5379007$ & $\times10^{-1}$ & $1.6$ & $\times10^{-5}$ &  & $0.011212$\\
4 & $1.7754713$ &  & $1.775428(54)$ &  & $1.3$ & $\times10^{-5}$ & $-9.5755901$ & $\times10^{-1}$ & $2.0$ & $\times10^{-5}$ &  & $0.017113$\\
5 & $4.4934150$ &  & $4.49328(37)$ &  & $1.5$ & $\times10^{-5}$ & $-2.8247860$ &  & $2.3$ & $\times10^{-5}$ &  & $0.026122$\\
6 & $1.2746739$ & $\times10^{1}$ & $1.27455(30)$ & $\times10^{1}$ & $1.8$ & $\times10^{-5}$ & $-8.9271143$ &  & $2.7$ & $\times10^{-5}$ &  & $0.038735$\\
7 & $3.9456159$ & $\times10^{1}$ & $3.9458(35)$ & $\times10^{1}$ & $2.0$ & $\times10^{-5}$ & $-2.9918529$ & $\times10^{1}$ & $3.3$ & $\times10^{-5}$ &  & $0.055301$\\
8 & $1.3122787$ & $\times10^{2}$ & $1.3097(19)$ & $\times10^{2}$ & $2.2$ & $\times10^{-5}$ & $-1.0563844$ & $\times10^{2}$ & $5.4$ & $\times10^{-5}$ &  & $0.075901$\\
9 & $4.6446232$ & $\times10^{2}$ & $4.630(18)$ & $\times10^{2}$ & $2.4$ & $\times10^{-5}$ & $-3.9126964$ & $\times10^{2}$ & $4.1$ & $\times10^{-5}$ &  & $0.100303$\\
10 & $1.7383950$ & $\times10^{3}$ & $1.714(10)$ & $\times10^{3}$ & $3.0$ & $\times10^{-5}$ & $-1.5159158$ & $\times10^{3}$ & $6.1$ & $\times10^{-5}$ &  & $0.127982$\\
11 & $6.8511259$ & $\times10^{3}$ & $6.93(17)$ & $\times10^{3}$ & $3.1$ & $\times10^{-5}$ & $-6.1315810$ & $\times10^{3}$ & $4.8$ & $\times10^{-5}$ &  & $0.158191$\\
12 & $2.8345123$ & $\times10^{4}$ & $2.800(71)$ & $\times10^{4}$ & $3.2$ & $\times10^{-5}$ & $-2.5861627$ & $\times10^{4}$ & $5.0$ & $\times10^{-5}$ &  & $0.190062$\\
13 & $1.2285038$ & $\times10^{5}$ & $1.208(54)$ & $\times10^{5}$ & $3.3$ & $\times10^{-5}$ & $-1.1363144$ & $\times10^{5}$ & $5.3$ & $\times10^{-5}$ &  & $0.222719$\\
14 & $5.5688264$ & $\times10^{5}$ & $5.00(16)$ & $\times10^{5}$ & $3.4$ & $\times10^{-5}$ & $-5.1991561$ & $\times10^{5}$ & $6.3$ & $\times10^{-5}$ &  & $0.255355$\\
15 & $2.6369442$ & $\times10^{6}$ & $2.54(19)$ & $\times10^{6}$ & $3.6$ & $\times10^{-5}$ & $-2.4762555$ & $\times10^{6}$ & $6.6$ & $\times10^{-5}$ &  & $0.287295$\\
16 & $1.3030422$ & $\times10^{7}$ & $1.24(15)$ & $\times10^{7}$ & $3.9$ & $\times10^{-5}$ & $-1.2273417$ & $\times10^{7}$ & $1.0$ & $\times10^{-4}$ &  & $0.318038$\\
17 & $6.7135540$ & $\times10^{7}$ & $4.72(15)$ & $\times10^{7}$ & $4.1$ & $\times10^{-5}$ & $-6.3273412$ & $\times10^{7}$ & $8.0$ & $\times10^{-5}$ &  & $0.347243$\\
18 & $3.6034541$ & $\times10^{8}$ & $3.41(62)$ & $\times10^{8}$ & $4.2$ & $\times10^{-5}$ & $-3.3936059$ & $\times10^{8}$ & $7.4$ & $\times10^{-5}$ &  & $0.374711$\\
19 & $2.0132906$ & $\times10^{9}$ & $2.6(1.3)$ & $\times10^{9}$ & $4.4$ & $\times10^{-5}$ & $-1.8920042$ & $\times10^{9}$ & $8.4$ & $\times10^{-5}$ &  & $0.400370$\\
20 & $1.1699650$ & $\times10^{10}$ & $9.6(1.8)$ & $\times10^{9}$ & $4.5$ & $\times10^{-5}$ & $-1.0961195$ & $\times10^{10}$ & $7.5$ & $\times10^{-5}$ & $0.54687(26)$ & $0.424249$\\
25 & $1.3567469$ & $\times10^{14}$ &  &  & $5.8$ & $\times10^{-5}$ & $-1.2455020$ & $\times10^{14}$ & $9.1$ & $\times10^{-5}$ & $0.51410(27)$ & $0.520475$\\
30 & $3.6820966$ & $\times10^{18}$ &  &  & $6.6$ & $\times10^{-5}$ & $-3.3166830$ & $\times10^{18}$ & $1.0$ & $\times10^{-4}$ & $0.48769(28)$ & $0.588902$\\
35 & $2.1033826$ & $\times10^{23}$ &  &  & $7.1$ & $\times10^{-5}$ & $-1.8671906$ & $\times10^{23}$ & $2.1$ & $\times10^{-4}$ & $0.46901(27)$ & $0.640109$\\
40 & $2.3125563$ & $\times10^{28}$ &  &  & $8.1$ & $\times10^{-5}$ & $-2.0300655$ & $\times10^{28}$ & $2.8$ & $\times10^{-4}$ & $0.45592(29)$ & $0.679974$\\
45 & $4.5506926$ & $\times10^{33}$ &  &  & $8.5$ & $\times10^{-5}$ & $-3.9596387$ & $\times10^{33}$ & $1.4$ & $\times10^{-4}$ & $0.44559(31)$ & $0.711920$\\
50 & $1.5109057$ & $\times10^{39}$ &  &  & $8.9$ & $\times10^{-5}$ & $-1.3053699$ & $\times10^{39}$ & $1.5$ & $\times10^{-4}$ & $0.43851(33)$ & $0.738088$\\
55 & $8.0627600$ & $\times10^{44}$ &  &  & $9.2$ & $\times10^{-5}$ & $-6.9254253$ & $\times10^{44}$ & $2.7$ & $\times10^{-4}$ & $0.43145(34)$ & $0.759906$\\
60 & $6.6409459$ & $\times10^{50}$ &  &  & $9.5$ & $\times10^{-5}$ & $-5.6777530$ & $\times10^{50}$ & $3.1$ & $\times10^{-4}$ & $0.42650(36)$ & $0.778380$\\
65 & $8.1580145$ & $\times10^{56}$ &  &  & $9.9$ & $\times10^{-5}$ & $-6.9462938$ & $\times10^{56}$ & $2.0$ & $\times10^{-4}$ & $0.42085(89)$ & $0.794193$\\
70 & $1.4515383$ & $\times10^{63}$ &  &  & $1.3$ & $\times10^{-4}$ & $-1.2323232$ & $\times10^{63}$ & $1.1$ & $\times10^{-3}$ & $0.41784(82)$ & $0.807890$\\
75 & $3.6476368$ & $\times10^{69}$ &  &  & $1.6$ & $\times10^{-4}$ & $-3.0832047$ & $\times10^{69}$ & $6.8$ & $\times10^{-4}$ & $0.41617(79)$ & $0.819977$\\
80 & $1.2658951$ & $\times10^{76}$ &  &  & $1.8$ & $\times10^{-4}$ & $-1.0684536$ & $\times10^{76}$ & $1.4$ & $\times10^{-3}$ & $0.41157(81)$ & $0.830524$\\
85 & $5.9512472$ & $\times10^{82}$ &  &  & $2.2$ & $\times10^{-4}$ & $-5.0116314$ & $\times10^{82}$ & $1.9$ & $\times10^{-3}$ & $0.40970(81)$ & $0.839923$\\
90 & $3.7253253$ & $\times10^{89}$ &  &  & $2.3$ & $\times10^{-4}$ & $-3.1295215$ & $\times10^{89}$ & $8.6$ & $\times10^{-4}$ & $0.40739(85)$ & $0.848415$\\
95 & $3.0570431$ & $\times10^{96}$ &  &  & $2.5$ & $\times10^{-4}$ & $-2.5629443$ & $\times10^{96}$ & $1.3$ & $\times10^{-3}$ & $0.40626(88)$ & $0.856024$\\
100 & $3.2434401$ & $\times10^{103}$ &  &  & $2.7$ & $\times10^{-4}$ & $-2.7137827$ & $\times10^{103}$ & $9.8$ & $\times10^{-4}$ & $0.40428(89)$ & $0.862881$\\
101 & $8.4934059$ & $\times10^{104}$ &  &  & $2.7$ & $\times10^{-4}$ & $-7.1009617$ & $\times10^{104}$ & $1.0$ & $\times10^{-3}$ &  & $0.864240$\\
110 & $7.516949$ & $\times10^{117}$ &  &  & $4.2$ & $\times10^{-4}$ & $-6.26219$ & $\times10^{117}$ & $2.5$ & $\times10^{-3}$ &  & $0.875039$\\
120 & $4.249680$ & $\times10^{132}$ &  &  & $5.4$ & $\times10^{-4}$ & $-3.54069$ & $\times10^{132}$ & $3.0$ & $\times10^{-3}$ &  & $0.885331$\\
130 & $5.446338$ & $\times10^{147}$ &  &  & $6.2$ & $\times10^{-4}$ & $-4.52393$ & $\times10^{147}$ & $3.1$ & $\times10^{-3}$ &  & $0.893801$\\
140 & $1.487803$ & $\times10^{163}$ &  &  & $7.8$ & $\times10^{-4}$ & $-1.22990$ & $\times10^{163}$ & $4.3$ & $\times10^{-3}$ &  & $0.901039$\\
150 & $8.215428$ & $\times10^{178}$ &  &  & $8.8$ & $\times10^{-4}$ & $-6.78650$ & $\times10^{178}$ & $4.6$ & $\times10^{-3}$ &  & $0.907397$\\
160 & $8.760058$ & $\times10^{194}$ &  &  & $1.3$ & $\times10^{-3}$ & $-7.22819$ & $\times10^{194}$ & $4.1$ & $\times10^{-2}$ &  & $0.913518$\\
170 & $1.730498$ & $\times10^{211}$ &  &  & $2.1$ & $\times10^{-3}$ & $-1.44144$ & $\times10^{211}$ & $9.7$ & $\times10^{-3}$ &  & $0.918696$\\
180 & $6.109169$ & $\times10^{227}$ &  &  & $2.6$ & $\times10^{-3}$ & $-5.04137$ & $\times10^{227}$ & $1.6$ & $\times10^{-2}$ &  & $0.922858$\\
190 & $3.735844$ & $\times10^{244}$ &  &  & $3.3$ & $\times10^{-3}$ & $-3.10095$ & $\times10^{244}$ & $2.4$ & $\times10^{-2}$ &  & $0.926725$\\
200 & $3.844913$ & $\times10^{261}$ &  &  & $4.4$ & $\times10^{-3}$ & $-3.20084$ & $\times10^{261}$ & $2.6$ & $\times10^{-2}$ &  & $0.930167$\\
210 & $6.491400$ & $\times10^{278}$ &  &  & $4.6$ & $\times10^{-3}$ & $-5.42414$ & $\times10^{278}$ & $2.6$ & $\times10^{-2}$ &  & $0.933341$\\
220 & $1.757734$ & $\times10^{296}$ &  &  & $5.0$ & $\times10^{-3}$ & $-1.45890$ & $\times10^{296}$ & $1.1$ & $\times10^{-2}$ &  & $0.937025$\\
230 & $7.457214$ & $\times10^{313}$ &  &  & $5.1$ & $\times10^{-3}$ & $-6.13554$ & $\times10^{313}$ & $1.8$ & $\times10^{-2}$ &  & $0.939240$\\
240 & $4.869811$ & $\times10^{331}$ &  &  & $5.3$ & $\times10^{-3}$ & $-3.99761$ & $\times10^{331}$ & $1.0$ & $\times10^{-2}$ &  & $0.941586$\\
250 & $4.802735$ & $\times10^{349}$ &  &  & $5.5$ & $\times10^{-3}$ & $-3.89739$ & $\times10^{349}$ & $1.8$ & $\times10^{-2}$ &  & $0.942917$\\
260 & $7.049558$ & $\times10^{367}$ &  &  & $5.6$ & $\times10^{-3}$ & $-5.81110$ & $\times10^{367}$ & $1.1$ & $\times10^{-2}$ &  & $0.944930$\\
270 & $1.515747$ & $\times10^{386}$ &  &  & $5.8$ & $\times10^{-3}$ & $-1.26885$ & $\times10^{386}$ & $2.9$ & $\times10^{-2}$ &  & $0.946926$\\
280 & $4.702604$ & $\times10^{404}$ &  &  & $5.9$ & $\times10^{-3}$ & $-3.91162$ & $\times10^{404}$ & $3.0$ & $\times10^{-2}$ &  & $0.947930$\\
290 & $2.083259$ & $\times10^{423}$ &  &  & $6.1$ & $\times10^{-3}$ & $-1.72252$ & $\times10^{423}$ & $2.4$ & $\times10^{-2}$ &  & $0.950370$\\
300 & $1.298937$ & $\times10^{442}$ &  &  & $6.6$ & $\times10^{-3}$ & $-1.06520$ & $\times10^{442}$ & $2.8$ & $\times10^{-2}$ &  & $0.952114$\\
310 & $1.127159$ & $\times10^{461}$ &  &  & $7.9$ & $\times10^{-3}$ & $-9.29327$ & $\times10^{460}$ & $4.2$ & $\times10^{-2}$ &  & $0.953247$\\
320 & $1.350431$ & $\times10^{480}$ &  &  & $8.4$ & $\times10^{-3}$ & $-1.09135$ & $\times10^{480}$ & $4.3$ & $\times10^{-2}$ &  & $0.956293$\\
325 & $5.253061$ & $\times10^{489}$ &  &  & $8.9$ & $\times10^{-3}$ & $-4.31071$ & $\times10^{489}$ & $9.7$ & $\times10^{-2}$ &  & $0.957490$\\
\bottomrule
\end{tabular}
\end{table}

\section{Perturbative expansion data}
\label{app:loopdata}

The exact one- and two-loop coefficients, $s_1=\sqrt\pi/2$, $d_1=-\sqrt\pi/24$, $u_1=\sqrt\pi/160$, $s_2=\pi/6$, $d_2=-91\pi/1944$ and $u_2=10537\pi/874800$, enter Table~\ref{tab:C:coeffs} as its first two rows. Table~\ref{tab:C:coeffs} lists the telescoped coefficients, whose relative errors grow from $9\times10^{-6}$ at $L=3$ to $2.7\times10^{-4}$ at $L=101$, $4.4\times10^{-3}$ at $L=200$, $6.6\times10^{-3}$ at $L=300$ and $8.9\times10^{-3}$ at $L=325$ for $s_L$, and stay below $2\times10^{-3}$ through $L=101$ and $10^{-1}$ through $L=325$ for $d_L$, together with $u_L/s_L$, known to $(5$--$22)\times10^{-4}$. For $L\le20$ the table also lists $s_L$ from the three-point coefficients of Ref.~\cite{Borinsky:2025stz}, converted by the Ward identity. They agree with ours within $2.5\sigma$ except at $L=14$ and 17, where they are lower by 10\% and 30\%, or $3.4\sigma$ and $13.6\sigma$ in their quoted errors. Independent direct runs at $L=120$, 150 and 200 give $R_L=0.8919(93)$, $0.9068(96)$ and $0.9078(229)$, within $0.7\sigma$, $0.1\sigma$ and $1.0\sigma$ of the telescoped values. The complete coefficients with their covariance will be provided as ancillary files.

The comparison with the calculated fluctuations was preregistered. The fit window $L=30$--64, the models and the test criteria were fixed before the high-statistics data were analysed, and the orders $L=65$--100 were kept for an out-of-sample check. We fit the ratios $q_L=R_{L+1}/R_L$, in which $C_s$ cancels, and for the slope $x_L=(d_{L+1}/s_{L+1})/(d_L/s_L)$, with the factorial form~\eqref{eq:B:RL}. The lower coefficients are fixed to their calculated values, the tested one is free, and the higher ones are free as a remainder. In the four-loop test the next coefficient, $c_4$, is constrained by a Gaussian prior from an auxiliary five-loop estimate, $c_4=29\pm47$. It comes from a Monte Carlo evaluation of the 979 five-loop vacuum graphs with the rules of App.~\ref{app:instanton}, with a statistical error of $\pm25$ and a margin of $\pm40$ from a comparison of two gauges. Table~\ref{tab:C:fits} reports the power-basis coefficients $c_k$, related to the fluctuation coefficients by Eq.~\eqref{eq:B:basis}. A loop is resolved if setting its fluctuation coefficient to zero, $\hat\kappa_2=0$ for three loops and $\hat\kappa_3=0$ for four, with the other free parameters refitted, costs $\Delta\chi^2\ge9$. It is compatible if in addition $|\text{pull}|\le2$, the fit has $p\ge0.01$, and the result is stable under $L_{\min}=35$, 40 and one more remainder term. On synthetic series these criteria accept the true coefficients in 98--99\% of cases.

Table~\ref{tab:C:fits} gives the results. The three-loop coefficient is resolved and compatible. With the $c_4$ prior the four-loop coefficient is also resolved and compatible, $c_3=-165.5\pm3.9$ with $\Delta\chi^2=119$. Without the prior a free fit gives only $c_3=-208\pm38$ and $\Delta\chi^2=5.0$, so this result is conditional on the five-loop estimate. The zero-parameter comparison resolves the four-loop term without it. With $\hat\kappa_1$--$\hat\kappa_3$ fixed and nothing fitted, the ratios give $\chi^2=57.4$ for the 51 orders $L=50$--100, against 190.6 with $\hat\kappa_3=0$. The difference comes from $L=50$--64, while at $L=65$--100 the four-loop term lies below the uncertainties and below $L\approx45$ the data need terms beyond four loops. The slope test is conditional because the influence functions of $x_L$ have Hill indices below 2.

The middle and bottom of Table~\ref{tab:C:fits} use no input from the saddle beyond the one listed. The action and the exponent $b$ come from fits of the growth of $s_L$ at $L\le325$, with the spread of acceptable forms and windows as a model error. Orders beyond $L=100$ sharpen the action fourfold and $b$ threefold, but not the fluctuation coefficients. The size comes from the limit $\rho_d$ of $d_L/s_L$. For an instanton $\rho_d=2-\mu_2/3$, Eq.~\eqref{eq:B:rho}, while on the diagram side $-d_L/s_L$ is the effective resistance of $\Sigma_G(0)$ averaged over all $L$-loop graphs, which measures the squared separation of the external vertices, $\langle x^2\rangle=6R_G$. Large-order Feynman graphs thus have the size of the instanton. The $p^4$ coefficient is extrapolated with the same rule. Since $u_L/s_L$ still falls from 0.547 at $L=20$ to 0.404 at $L=100$, 99\% of the variance of $\rho_4$ comes from the extrapolation, and the control $d_L/s_L$, analysed in the same way, reproduces $\rho_d$. With the instanton $\mu_2$, Eq.~\eqref{eq:B:rho} turns $\rho_4$ into $\mu_4=181.36\pm0.56$, against 181.39608.

\begin{table}[t]
\caption{\label{tab:C:fits}Instanton data from the coefficients. Top, comparison with the calculated fluctuations,
with intervals from 1.5 times the per-order covariance and $\Delta\chi^2$ for omitting the loop order. Middle and bottom,
data-only extraction and the $p^4$ measurement, with total (statistical and model) uncertainties.}
\begin{ruledtabular}
\begin{tabular}{lllll}
quantity & fit inputs & instanton & perturbation & remarks\\
\hline
$c_2$ (3 loops) & $c_1$ fixed; $c_2,c_3,c_4$ free & $66.9994(5)$ & $65.34\pm1.31$ & pull $-1.27$, $\Delta\chi^2=1123$, $\chi^2=19.2/32$\\
$c_3$ (4 loops) & $c_1,c_2$ fixed; $c_3,c_4,c_5$ free, $c_4$ prior & $-171.37(40)$ & $-165.5\pm3.9$ & pull $+1.49$, $\Delta\chi^2=119$, $\chi^2=19.6/33$\\
$\hat\kappa_3$ (4 loops) & all fixed, $L=50$--100 & $-48.15(40)$ &  & $\chi^2=57.4/51$; 190.6 if $\hat\kappa_3=0$\\
$e_2$ (slope) & $e_1$ fixed; $e_2,e_3,e_4$ free, $e_3$ prior & $8.72(15)$ & $9.11\pm0.36$ & pull $+1.00$, $\Delta\chi^2=654$; conditional\\
\hline
$S_0=(4\pi)^{3/2}/a$ & none & $174.6409$ & $174.650\pm0.057$ & statistical error 0.005\\
$b$ & none & $5/2$ & $2.51\pm0.10$ & same fit\\
$S_0$ & $b$ & $174.6409$ & $174.635\pm0.019$ & \\
$C_s$ & $a$, $b$ & $85.1315$ & $84.99\pm0.30$ & \\
$\mu_2=3(2-\rho_d)$ & none & $8.43506$ & $8.427\pm0.012$ & $\rho_d=-0.8090\pm0.0041$\\
$c_1$ & $a$, $b$, $C_s$ & $-14.366$ & $-14.40\pm0.20$ & \\
\hline
$\rho_4$ & none & $0.376290$ & $0.37560\pm0.00939$ & pull $-0.07$; statistical error 0.00078\\
$\rho_d$ (control) & none & $-0.811688$ & $-0.80896\pm0.00381$ & pull $+0.72$\\
\end{tabular}
\end{ruledtabular}
\end{table}

\end{document}